\documentclass[12pt]{article}

\usepackage[T1]{fontenc}
\usepackage[utf8]{inputenc}
\usepackage[svgnames,table]{xcolor}
\usepackage[unicode=true]{hyperref} 
\hypersetup{
colorlinks=true,
  linkcolor={blue},
  citecolor={Blue},
  urlcolor={Blue}}
  
\usepackage{amsmath,amssymb}
\usepackage{lmodern}
\usepackage{graphicx}
\usepackage{natbib}
\usepackage{bookmark}
\usepackage{float}
\usepackage{caption}
\usepackage{subcaption}
\usepackage[margin=1in]{geometry}
\usepackage{mathtools}
\usepackage{pifont}
\usepackage{threeparttable}
\usepackage{multirow}
\usepackage{orcidlink}
\newcommand{\pr}{\mathcal{P}}
\newcommand{\m}{\mathbf{m}}
\newcommand{\M}{\mathbf{M}}
\newcommand{\x}{\mathbf{x}}
\newcommand{\X}{\mathbf{X}}
\newcommand{\Z}{\mathbf{Z}}
\newcommand{\z}{\mathbf{z}}
\newcommand{\btheta}{\boldsymbol{\theta}}

\newcommand{\boldeta}{\boldsymbol{\eta}}
\newcommand{\cmark}{\ding{51}}
\newcommand{\xmark}{\ding{55}}
\newcommand{\Ev}{\boldsymbol{\mathcal{E}}}
\newcommand{\ev}{\boldsymbol{\varepsilon}}
\newcommand{\Tr}{\mathcal{T}}
\newcommand{\Data}{\mathcal{D}}
\newcommand{\Risk}{\mathcal{R}}
\newcommand{\E}{\mathbb{E}}
\newcommand{\Score}{\mathcal{S}}
\newcommand{\mx}{\mathbf{m}_{\mathbf{X}}}

\newcommand{\Mx}{\mathbf{M}_{\mathbf{X}}}
\newcommand{\ArticleTitle}{Prediction with Missing Data: Target Probabilities and Missingness Mechanisms}
\usepackage{tikz}
\usetikzlibrary{shapes,
decorations,
arrows,
calc,
arrows.meta,
fit,
positioning}
\tikzset{-Latex,semithick}

\begin{document}

\title{\bf \ArticleTitle}
  \author{Pierre Catoire,\orcidlink{0000-0002-1849-9781}\thanks{
    The authors gratefully acknowledge Pr. Cédric Gil-Jardiné, Dr Cécile Moreau and the Emergency Department of the University Hospital of Bordeaux (France) for their contribution in collecting application data.}\\
    Robin Genuer,\orcidlink{0000-0002-0981-3943}\\
    Cécile Proust-Lima\orcidlink{0000-0002-9884-955X}\\
    Univ. Bordeaux, INSERM, BPH, U1219\\F-33000 Bordeaux, France}

  \date{18 March 2026}
  \maketitle

\bigskip
\begin{abstract}
Conditions ensuring optimal parameter estimation in the presence of missing data are well established in inference, typically relying on the Missing-at-Random (MAR) assumption. In prediction, similar principles are often assumed to apply. However, methods considered biased in inference, such as pattern sub-modelling or unconditional imputation, have been shown to achieve optimal predictive performance under any missingness mechanism, including non-MAR (MNAR). To explain this apparent contradiction, we introduce a new formal framework for describing missingness in prediction. Central to this framework is a distinction between two prediction targets, defined according to whether or not the indicator of observation of the predictors is exploited to predict the outcome. This distinction leads to a classification of the missingness mechanisms describing the conditions under which these targets are equal, and when consistent prediction of each is achievable. A key result is that both targets may be consistently predicted under conditions weaker than MAR. We discuss the implications of this paradigm for handling missing data in prediction, distinguishing between missingness at development, validation and deployment of a forecaster. The findings are illustrated using simulated data and a real-world application with the prediction of significant injury after trauma upon arrival at the emergency department. 

\end{abstract}

\vfill

\newpage
\renewcommand{\baselinestretch}{1.8}
\small\normalsize

\section{Introduction}\label{sec:intro}

Missing data are ubiquitous in statistical analyses and must be handled appropriately. In inferential settings, the objective is to estimate parameters characterising the relationship between variables. When the probability of observing the variables depends on unobserved values, the mechanism is termed "Missing Not at Random" (MNAR), and standard procedures such as Maximum Likelihood Estimation (MLE) or Multiple Imputation (MI) generally fail to provide unbiased parameter estimates. Consistent estimation with such methods \citep{rubin_inference_1976, rubin_multiple_1987} requires the more restrictive "Missing at Random" (MAR) assumption under which the probability of observing a variable is independent of its unobserved value given the observations. 

In prediction, the goal is different: rather than estimating interpretable parameters, one seeks an algorithm that provides an accurate estimate of the outcome given available information. Recent results have suggested that methods considered as biased in inference, such as pattern sub-modelling or unconditional imputation, may actually provide consistent prediction under MNAR mechanisms \citep{fletcher_mercaldo_missing_2020, le_morvan_whats_2021, josse_consistency_2024}.

When defining a prediction algorithm, what is meant by available information is critical. It may refer only to the values of the observed predictors, or it may include both these values and the information that they have been observed, i.e. the "observation pattern" of the predictors. When observation or missingness of the predictors carries information about the outcome, the predictions based on these two sets differ. Consider a model predicting severe traumatic brain injury ($\text{TBI}$) from anticoagulant use ($\text{Med}$). Here, observing $\text{Med}$ ($M_{\text{Med}}$, the missingness indicator of $\text{Med}$) may depend on whether or not the patient has a brain injury, and is therefore able to report her/his medications. Here, observing $\text{Med}$ makes the brain injury less likely, and missing it makes a brain injury more likely. Without considering the information carried by the observation or missingness of the predictor, the probability of brain injury is $\Pr(\text{TBI} \mid \text{Med})$ if $\text{Med}$ is observed, and $\Pr(\text{TBI})$ if $\text{Med}$ is missing. This corresponds to the probability of the outcome conditionally on observed predictors, which we refer to as "Missingness Unconditional" (MU) probability distribution. Including $M_{\text{Med}}$ in the information used for prediction, the probability of brain injury is evaluated by $\Pr(\text{TBI} \mid \text{Med}, M_{\text{Med}} = 0)$ if $\text{Med}$ is observed, and $\Pr(\text{TBI} \mid M_{\text{Med}} = 1)$ if $\text{Med}$ is missing. This corresponds to the probability distribution of the outcome conditionally on the observed predictors and the observation pattern, which we refer to as "Missingness Conditional" (MC) probability distribution (see section \ref{sec:sub:what-is-to-predict}).

The MU/MC distinction provides a possible explanation to the apparent tension in the literature between MAR being a necessary condition for consistent inference, but not necessary for consistent prediction. The key insight is that consistent MC prediction can be achieved under any missingness mechanism of the predictors, whereas MLE and MI target the MU probability distribution which may not be predictable under MNAR.

A second distinction between inference and prediction is that missingness may occur at different stages: when the algorithm is built (development), when its predictive performance is assessed (validation), and when it is ultimately applied on a new individual (deployment). While MLE may be performed at development with missing data using algorithm such as Expectation-Maximisation \citep{dempster_maximum_1977}, and MI may be applied both at development and validation, they cannot generally be applied at deployment when predictions must be produced for a single individual with missing predictors. Dedicated procedures are therefore required, and their theoretical properties depend on the chosen prediction target.

In this work, we formalise the distinction between MU and MC prediction targets and propose a nomenclature of missingness mechanisms tailored to prediction problems. We characterise the assumptions under which consistent prediction of each target is achievable, with the notable result that both targets may be estimated under conditions weaker than MAR. We further review available procedures for handling missing predictors at development, validation and deployment, and illustrate their behaviour in simulation studies under various mechanisms, and apply the framework to real-world data for prediction of severe trauma in the emergency department.

\section{Notations and definitions for inference and prediction}\label{sec-definitions}

\subsection{Initial notations}\label{sec:sub:initial-notations}

We denote random variables by uppercase characters, their realisations by lowercase characters, and  vectors with boldfaced characters. The symbol $\perp$ is used for conditional independence, i.e. $A \perp B \mid C$ means that $A$ is conditionally independent from $B$ given $C$ for all values of $A$, $B$ and $C$. We write $A=a \perp B \mid C$ to denote a value-specific (or context-specific) conditional independence \citep{boutilier_context_1996}, meaning that $\Pr(A =a \mid B = b, C = c) = \Pr(A=a \mid C = c)$ for all values of $b$ and $c$ such that $\Pr(B=b, C=c) > 0$.

We consider a set of $l$ variables $\Z = \{Z_1, \ldots, Z_l\}$, with each variable defined over a domain $\mathcal{Z}_k, k \in \{1, \ldots, l\}$. We note $\z$ a realisation of $\Z$ with $\z = \{z_1, \ldots, z_l\} \in \mathbf{\mathcal{Z}}=\mathcal{Z}_1 \times \ldots \times \mathcal{Z}_l$. We note $\M_\Z = \{M_1, \ldots, M_l\}$ the set of missingness indicators, each of them defined over $\{0, 1\}$ with $M_k = 0$ if $Z_k$ is observed, and $M_k = 1$ if $Z_k$ is missing. The realised values of $\M_\Z$ are noted $\m_\Z \in \{0,1\}^l$, and called the "observation pattern".

For any pattern $\m_\Z$, we define a partition of $\Z$ between observed and missing subsets of variables
\begin{equation*}
    \begin{cases}
        \Z_o(\m_\Z) = \{Z_k, k \in \{1, \ldots, l\} \mbox{ such that } M_k = 0\} \\
        \Z_m(\m_\Z) = \{Z_k, k \in \{1, \ldots, l\} \mbox{ such that }  M_k = 1\},
    \end{cases}
\end{equation*}

and their realisations are noted $\z_o(\m_\Z)$ and $\z_m(\m_\Z)$, respectively. When non-ambiguous, $\Z_o(\m_\Z)$ and $\z_o(\m_\Z)$ are simplified to $\Z_o$ and $\z_o$, and similarly for $\Z_m$ and $\z_m$.

In many scenarios including prediction, the variables are split into independent variables (or predictors in prediction context) $\X = \{X_1, \ldots, X_p\}$ and a dependent variable (or outcome) $Y$. Domains are noted $\boldsymbol{\mathcal{X}} = \mathcal{X}_1 \times \ldots \times \mathcal{X}_p$ and $\mathcal{Y}$. We note $\Mx$ and $M_Y$ their missingness indicators and $\mx$ and $m_Y$ their realisations. We also note $\X_o(\mx)$ and $\X_m(\mx)$ the observed and missing variables for an observation pattern $\mx$, with realisations $\x_o(\mx)$ and $\x_m(\mx)$, simplified as $\X_o$, $\X_m$, $\x_o$ and $\x_m$ when non ambiguous. We assume that $\X, Y, \Mx, M_Y$ admit a probability distribution $\Pr(\X, Y, \Mx, M_Y)$.

\subsection{Missing values in inference}\label{sec:sub:missing_values-in-inference}

Inference requires modelling the relation between $Y$ and $\X$ with a function $f(\x; \btheta)$, where $\btheta$ is a vector of parameters. Given a finite sample of $N$ observations of $Y, \X$ written $\Data^N = \{y_i, \x_i\}_{i=1}^N$, inference is performed using an estimation procedure $\mathcal{I}\left(\Data^N\right) = \hat{\btheta}^N$, with $\hat{\btheta}^N$ the estimated parameters. The "optimality" of the estimation procedure in inference can be defined either as unbiasedness ($\E\left[\hat{\btheta}^N\right] = \btheta$) or consistency ($\hat{\btheta}^N \rightarrow \btheta$). Maximum Likelihood Estimation (MLE) achieves both asymptotically unbiasedness and consistency on complete data if the model is well-specified \citep{vandervaart_asymptotic_1998}.

In presence of missing data, these properties can be achieved only under certain conditions. Notably, a key result from \cite{rubin_inference_1976} defines the "Missing at Random" (MAR) assumption as the independence between observation of variables and missing variables, conditionally on the observed variables. In its original formulation, MAR is stated as $\M_\Z=\m_\Z \perp \Z_m(\m_\Z) \mid \Z_o(\m_\Z)$ for all $\m_Z$ with strictly positive probability. However, when $Y$ and $\X$ are distinguished, this definition must be separated for cases with observed and missing $Y$:
\begin{equation}
    \begin{cases}
        \Mx = \mx, M_Y = 0 \perp \X_m(\mx) \mid \X_o(\mx), Y \\
        \Mx = \mx, M_Y = 1 \perp \X_m(\mx), Y \mid \X_o(\mx)
    \end{cases}
    \tag{MAR with variables $Y,\X$}
\end{equation}

MAR guarantees that MLE over observed data $\Data_o^N = \{y_{o,i},\x_{o,i}\}_{i=1}^N$ achieves asymptotically unbiased and consistent parameters estimates \citep{little_statistical_1987}, for instance using an Expectation-Maximisation (EM) algorithm \citep{dempster_maximum_1977}. Another common procedure is to impute multiple plausible (Multiple Imputation, MI) values of the missing predictors, using imputation functions estimated on observed data, generating multiple datasets on which estimation is performed, before pooling the parameters obtained on each imputed dataset. This procedure has also been proven to be asymptotically unbiased and consistent \citep{rubin_multiple_1987} if the prediction and imputation models are well-specified.

A stronger missingness assumption, called "Missing Completely at Random" (MCAR), is when the observation of variables is independent of both the observed and missing variables:
\begin{equation}
\Mx, M_Y \perp \X, Y
\tag{MCAR}
\end{equation}

Note that MCAR implies MAR. When MAR does not hold, the missingness mechanism is said "Missing Not at Random" (MNAR). In such cases, unbiased inference is not achievable using MLE or MI on observed data. Explicit modelling of the missingness mechanism is required, the validity of which being impossible to assess in practice \citep{little_test_1988}.

\subsection{Missing values in prediction}\label{sec:sub:missing-values-in-prediction}

\subsubsection{Forecasters and training procedures}\label{sec:subsub:forecasters-and-training-procedures}

Prediction aims to construct a forecaster that estimates the probability of the outcome given available evidence. In this article, prediction is framed as the estimation of the full conditional distribution of the outcome, rather than a function returning a point estimate (such as the conditional mean in regression or the most likely class in classification). Indeed, point predictions can be viewed as functionals of the conditional distribution. Therefore, consistency for the full conditional distribution implies consistency of derived regression and classification rules. Moreover, in many applied settings, including medical decision-making, the entire predicted distribution of the outcome is relevant. For example, probabilities of competing diagnoses may be combined with clinical outcome utilities \citep{gneiting_probabilistic_2014}.

We denote by "evidence" $\Ev$ the information set exploited by the forecaster, with $\ev$ its realised values. The content of $\Ev$ depends on the deployment setting and whether or not the information carried by the observation pattern is exploited. First, if the forecaster requires all predictors to be observed, $\Ev$ contains the set of all predictors $\X$. If the forecaster is deployed on cases with missing predictors, $\Ev$ contains the subset $\X_o$ of observed predictors. Second, the forecaster may exploit the information carried by the observation pattern. In such case, $\Ev$ also contains $\M_\X$. The implications of including or not the observation pattern in the evidence exploited to predict the outcome are central in this work, and are developed in section \ref{sec:nomenclature}. Without loss of generality, we therefore define the evidence $\Ev$ as a subset of $(\X, \M_\X)$ and specify its precise content for each prediction setting. A forecaster is thus a function $\pr$ such that $\pr (\ev) \in \mathbb{P}_Y$, with $\mathbb{P}_\mathcal{Y}$ the set of probability distributions for $\mathcal{Y}$.

We define a training procedure $\Tr$ as a function returning a forecaster $\hat{\pr}^N$ from a finite independent and identically distributed (i.i.d.) sample of observations $\Data_o^N$:
\begin{equation}
    \Tr(\Data_o^N) = \hat{\pr}^N \in \mathcal{H}_{\Ev} \tag{Training procedure}
\end{equation}

with $\mathcal{H}_{\Ev}$ a family of forecasters exploiting the evidence set $\Ev$. Forecasters may be constructed using parametric models (e.g., generalised linear models), where the training procedure comprises an inference procedure $\mathcal{I}$ on the parametric model, and the computation of the corresponding predicted probability distribution of the outcome. Non-parametric approaches (e.g., tree-based or kernel methods) still yield interpretable probabilities, despite the absence of interpretable parameters. In parametric settings, the consistency of the prediction holds if the estimation procedure $\mathcal{I}$ is itself consistent and under correct model specification.

\subsubsection{Optimality in prediction}\label{sec:subsub:optimality-in-prediction}

The "optimality" in prediction is usually defined in terms of risk $\Risk$. For point estimates, the risk is defined as the expected loss, using loss functions $\mathcal{L} \colon \mathcal{Y} \times \mathcal{Y} \mapsto \mathbb{R}$. A common choice is the Squared Error loss yielding the Mean Squared Error (MSE) risk. For forecasters estimating the conditional distribution, the risk is defined as the expected score $\Risk(\pr) = \E\left[\Score(Y,\pr(\ev))\right]$, with $\Score \colon \mathcal{Y} \times \mathbb{P}_\mathcal{Y} \mapsto \mathbb{R}$ a scoring rule \citep{gneiting_strictly_2007}. 

The lowest achievable risk within a family of forecasters $\mathcal{H}_{\Ev}$ is called the "Bayes risk" $\Risk^* = \underset{\pr \in \mathcal{H}_{\Ev}}{\inf} \ \Risk(\pr)$, and a Bayes optimal forecaster is a forecaster $\pr^*$ that achieves the Bayes risk, i.e. $\Risk(\pr^*) = \Risk^*$ \citep{lehmann_theory_1998}. The Bayes optimal forecaster is the true conditional distribution $\Pr(Y \mid \Ev)$ if this distribution belongs to the class $\mathcal{H}_{\Ev}$ and if the scoring rule is strictly proper, such as Brier score or log score \citep{gneiting_probabilistic_2014}.

Because the forecaster $\Tr(\Data^N)$ depends on the random sample $\Data^N$, its risk $\Risk(\Tr(\Data^N))$ is itself a random variable. 
We define therefore a training procedure $\Tr$ as "Bayes consistent" if:
\begin{equation}
\Risk\left(\Tr(\Data^N)\right)
\xrightarrow{\textrm{P}}
\Risk^*
\qquad \text{as } N \to \infty ,
\tag{Bayes consistency}
\end{equation}

where the convergence in probability is taken with respect to the randomness of the training sample $\Data^N$.

\subsubsection{Role of the MAR assumption in prediction}\label{sec:subsub:role-of-the-mar-assumption-in-prediction}

When missing values are present at development, and if a parametric forecaster with a well-specified model is considered, the MAR assumption guarantees consistent estimation of the parameters, and hence the training procedure is Bayes consistent. However, several authors \citep{le_morvan_whats_2021, bertsimas_simple_2024, josse_consistency_2024} have highlighted that, contrary to inference, the MAR assumption is not a necessary condition for consistent prediction. In particular, \cite{le_morvan_whats_2021} have shown that imputing missing predictors deterministically from a function of observed predictors, and training the forecaster on the imputed values (impute-then-regress procedure), is Bayes consistent for almost all imputation functions. Notably, this result holds under any missingness mechanism of the predictors including MNAR, which contrasts with results known in inference. Similarly, \cite{fletcher_mercaldo_missing_2020} have shown that the "Pattern Submodel" (PS) procedure, consisting of training one sub-model per observation pattern, achieves Bayes consistency for any missingness mechanism of the predictors, including MNAR, and for all observation patterns (see section \ref{sec:methods}). From this, MAR appears as a sufficient condition (for parametric forecasters) but not necessary for consistent prediction, contrasting with inference context.

\section{A nomenclature for probability distribution targets and missingness mechanisms for prediction}\label{sec:nomenclature}

In this section, we introduce a novel nomenclature offering an explanation to the apparent conflict between the above results on missingness in prediction and the established results from inference.

\subsection{What is to predict? Missingness-Unconditional and Missingness-Conditional probabilities}\label{sec:sub:what-is-to-predict}

When predicting the outcome, the evidence available to perform that prediction can be understood either as the value of the observed predictors only $\X_o$, or as the value  of the observed predictors and of the observation pattern $\Mx$. Indeed, as illustrated in the example of the introduction, the fact that some predictors are observed or missing may carry information on the outcome which is not provided by the value of the observed predictors. In such a situation, two quantities may be sought to be estimated by a forecaster which we call the "Missingness-Unconditional" (MU) probability distribution, and the "Missingness-Conditional" (MC) probability distribution:
\begin{equation*}
\begin{aligned}
    \textrm{MU} \coloneqq \Pr(Y \mid \X_o = \x_o) \\
    \textrm{MC} \coloneqq \Pr(Y \mid \X_o = \x_o, \M_\X = \m_\X)
\end{aligned}
    \tag{MU and MC probability distributions}
\end{equation*}

This distinction highlights the possible ambiguity when determining whether a forecaster is "optimal"  or not. If MU and MC are not equal, a forecaster may, at best, provide an optimal prediction of only one of the two distributions. The assumptions used to describe missingness mechanisms in prediction should therefore clarify whether MU and MC are equal, and under which conditions consistent prediction of both (if they are equal) or each separately (if they differ) is essential for designing the analysis. In the following, we propose a novel nomenclature of missingness mechanisms in prediction, discuss under which of these assumptions consistent prediction may be achieved, show the implication relations between them and with the MAR and MCAR assumptions used in inference, and illustrate the possible combinations of assumptions with examples.

\subsection{Nomenclature of missingness mechanisms for prediction}\label{sec:sub:nomenclature-of-missingness-mechanisms-for-prediction}

\subsubsection{Informative missingness of the predictors for the outcome}\label{sec:subsub:informative-missingness-of-the-predictors-for-the-outcome}

We define the missingness mechanism of the predictors as "Non-Informative Missingness for the Outcome" (NIMO) if for any observation pattern, the outcome is conditionally independent of the observation pattern given the value of the observed predictors for that pattern:
\begin{equation*}
Y \perp \Mx = \mx \mid \X_o(\mx) \qquad \forall \ \mx \colon \Pr(\Mx = \mx) > 0
    \tag{NIMO}
\end{equation*}

Under NIMO, MU and MC coincide for all patterns with positive probability. Note that this distinction is also relevant when all predictors are observed: in the introductory example, the fact that the patient is able to report the use of anticoagulant medication, i.e. $M_\mathrm{Med} = 0$, is suggestive of absence of traumatic brain injury. If the forecaster is intended to be applied only on individuals where all predictors are expected to be observed, a weaker version of the NIMO assumption may be relevant, requesting conditional independence only for the pattern $\mx = \{0\}^p$ where all predictors are observed (and therefore $\X_o = \X$, assuming $\Pr(\Mx = \{0\}^p) > 0$). This assumption is called "Non-Informative Complete Observation of the predictors" (NICO):
\begin{equation}
Y \perp \Mx = \{0\}^p \mid \X
    \tag{NICO}
\end{equation}

"Informative Missingness for the Outcome" (IMO) and "Informative Complete Observation of the predictors for the Outcome" (ICO) are the situations where NIMO and NICO do not hold, respectively.

\subsubsection{Missingness at Random for the predictors}\label{sec:subsub:missingness-at-random-for-the-predictors}

Two other assumptions which describe the conditional independence relations between the observation pattern of the predictors and the missing predictors help determine whether consistent prediction can be achieved or not. We discussed in section \ref{sec:sub:missing_values-in-inference} that MAR applied in contexts where outcome and predictors are distinct implies distinguishing between cases with missing or observed outcome. By considering only the relationship between variables and observation pattern of the predictors, we can define "Missingness at Random for the predictors" (MARX) when the outcome is missing (MARX-YM) or observed (MARX-YO):
\begin{equation*}
    \Mx \perp \X_m, Y \mid \X_o \quad \textrm{(MARX-YM)} \qquad \Mx \perp \X_m \mid \X_o, Y \quad \textrm{(MARX-YO)}
\end{equation*}

\subsection{Sufficient conditions for Bayes consistent procedures}\label{sec:sub:sufficient-conditions-for-Bayes consistent-procedures}

Bayes consistency of training procedures can be reconsidered with regard to the MU/MC distinction. Indeed, as the Bayes risk is defined as the lowest achievable risk among a class of forecasters, if the class of forecasters is $\mathcal{H}_{\X_o}$ (the set of functions with evidence $\Ev = \X_o$), then the optimal predictor is the MU probability distribution which achieves the MU-Bayes risk $\Risk^*_\textrm{MU}$, whereas if the class of forecasters is $\mathcal{H}_{\X_o, \Mx}$ (the set of functions with evidence $\Ev= \{ \X_o, \Mx \}$), then the optimal predictor is the MC probability distribution, achieving the MC-Bayes risk $\Risk^*_{\textrm{MC}}$:
\begin{equation*}
\Risk^*_{\textrm{MU}} = \underset{\pr \in \mathcal{H}_{\X_o}}{\inf}\Risk(\pr) \quad ; \quad \Risk^*_{\textrm{MC}} = \underset{\pr \in \mathcal{H}_{\X_o, \Mx}}{\inf}\Risk(\pr)\tag{MU and MC Bayes risks}
\end{equation*}

We describe in the following conditions sufficient to achieve Bayes consistency for both risks, assuming that MU and MC belong to $\mathcal{H}_{\X_o}$ and $\mathcal{H}_{\X_o, \Mx}$, respectively.

\subsubsection{Bayes consistent training procedures for MU risk}\label{sec:subsub:Bayes-consistent-training-procedures-for-mu-risk}

If the evidence includes the observed predictors only and if the MU probability distribution belongs to $\mathcal{H}_{\X_o}$, then the Bayes optimal forecaster is the MU probability distribution:
\begin{equation*}
    \textrm{MU} = \underset{\pr \in \mathcal{H}_{\X_o}}{\arg \min}\ \Risk(\pr)
\end{equation*}

As MLE and MI are consistent parameter estimation procedures under MAR when the model $f(\x; \btheta)$ (and the imputation models for MI) is well-specified, they are also Bayes consistent training procedures. This makes MAR a sufficient condition for consistent prediction. However, we can show that MLE and MI are Bayes consistent for MU risk under conditions weaker than MAR. Notably, under MARX-YO  with additional assumptions on the conditional independence of $M_Y$, a MLE or MI training procedure $\Tr$ applied on the subset of the training data where $Y$ is observed (i.e. $M_Y = 0$) noted $\Data^{N_0}_{M_Y=0}$ is MU-Bayes consistent. Formally,
\begin{equation*}
\begin{rcases}
\text{MARX-YO} \\
M_Y \perp \Mx \mid \X_o, \X_m, Y \\
M_Y \perp Y \mid \X_o
\end{rcases} \Rightarrow \Risk\left(\Tr(\Data^{N_0}_{M_Y=0}) \right)\rightarrow \Risk^*_\textrm{MU}
\end{equation*}

The demonstration is provided in Supplementary Material \ref{sec:sub:consistency-MLE-MARXYO}.

\subsubsection{Bayes consistent training procedures for MC risk}\label{sec:subsub:Bayes-consistent-training-procedures-for-mc-risk}

If the evidence includes the observed predictors and the observation pattern and if the MC probability distribution belongs to $\mathcal{H}_{\X_o, \Mx}$, then the Bayes optimal forecaster is the MC probability distribution:
\begin{equation*}
    \textrm{MC} = \underset{\pr \in \mathcal{H}_{\X_o, \Mx}}{\arg \min}\ \Risk(\pr)
\end{equation*}

\cite{le_morvan_whats_2021} and \cite{fletcher_mercaldo_missing_2020} showed that impute-then-regress procedures within almost all imputation functions and Pattern Sub-models both achieve MC-Bayes consistency under any missingness mechanism of the predictors, including MNAR. We present in section \ref{sec:methods} other training procedures which achieve MC-Bayes consistency.

As an illustrative example, we consider the consistency of Complete Case Analysis (CCA), where all observations with missing predictors are excluded from the development set. Here, the development set is indeed the subsample of the original set where $\m_{i,\X} = \{0\}^p$ and $\x_i, y_i \sim \Pr(\X, Y \mid \Mx = \{0\}^p)$ for every $i \in \{1, \ldots, N_{X_0}\}$ (with $N_{X_0}$ the number of cases with all predictors observed). As no missing value is remaining in this subsample, MLE yields parameter estimates converging to $\btheta_{\Mx = \{0\}^p}$ under correct specification of the model $f(\x; \btheta_{\Mx = \{0\}^p}) = \Pr(Y \mid \X = \x, \Mx = \{0\}^p)$. When the model is deployed on a new individual $j$ with all predictors observed $\x_j$, the observation pattern for this individual is $\m_{j,\X} = \{0\}^p$. The prediction obtained by computing $f(\x_j; \hat{\btheta}_{\Mx = \{0\}^p})$ is therefore asymptotically MC-Bayes consistent, without assumption about the missingness mechanism. This contrasts with results known from inference where CCA is unbiased only under MCAR, because inference targets parameter $\btheta$ of $\Pr(Y \mid \X)$.

Pattern Sub-models can be seen as a generalisation of CCA where a model for each observation pattern is estimated, explaining the result obtained by \cite{fletcher_mercaldo_missing_2020}. Similarly, results obtained by \cite{le_morvan_whats_2021} can be interpreted as achieving MC-Bayes consistency, given the fact that almost all imputation functions map the observations into pattern-specific subsets of the $\mathcal{X} \times \mathcal{Y}$ space, which can be differentiated by a flexible enough training procedure.

\subsection{Identifiability of MU and MC prediction targets}\label{sec:sub:achievability-of-consistent-MU-and-MC-prediction}

These results suggest that consistent MU prediction generally requires stronger assumptions than MC prediction. The reason may be that predictors are always observed jointly with their observation pattern, so that the conditional distribution of $(Y, \X_o(\mx) \mid \Mx = \mx, M_Y = 0)$ is identifiable from the observed data. From this, MC probability distribution can be estimated directly from observed data. 

In contrast, estimating the MU probability distribution requires to estimate the conditional probability of $Y$ given the predictors, without conditioning on the observation pattern. Consider a new individual $j$ with an observation pattern $\m_{\X, j}$ and observed predictors $\x_{o,j}(\m_{\X,j})$. We can decompose the MU probability distribution for this individual as a marginalisation over all observation patterns, that is:

\begin{equation*}
    \Pr \left ( Y \mid \X_o(\m_{\X,j}) = \x_{o,j}(\m_{\X,j}) \right ) = 
    \underset{\m_\X}{\sum} \Pr \left ( Y, \M_\X = \m_\X \mid \X_o(\m_{\X,j}) = \x_{o,j}(\m_{\X,j}) \right )
\end{equation*}

Importantly, this marginalisation is performed over all possible observation patterns, and notably over patterns different from $\m_{\X,j}$. For instance, consider predicting $Y$ from the predictors $X_1$ (potentially missing) and $X_2$ (always observed), with missingness indicator $M_1$. For $m_{1,j}= 0$ and $\x_{o,j}(m_{1,j}) = (x_{1,j}, x_{2,j})$, the marginalisation above yields:

\begin{align*}
    \Pr(Y \mid X_1 = x_{1,j}, X_2 = x_{2,j}) &= \Pr(Y, M_1 = 0 \mid X_1 = x_{1,j}, X_2 = x_{2,j}) \\
    &+ \Pr(Y, M_1 = 1 \mid X_1 = x_{1,j}, X_2 = x_{2,j})
\end{align*}

But $\Pr(Y, M_1 = 1 \mid X_1 = x_{1,j}, X_2 = x_{2,j})$ cannot be estimated from observed data alone without additional assumptions, as $M_1 = 1$ and $X_1$ are not observed jointly.

\subsection{Choosing between MC and MU targets and missingness shift}\label{sec:sub:choice-between-mc-and-mu-targets-and-missingness-shift}

Which probability distribution to target between MU and MC is subject to debate. On the one hand, MC probability distribution uses more information than MU, implying a lower Bayes risk and therefore a more accurate prediction. On the other hand, the additional information carried by the observation pattern of the predictors may be distorted if the missingness mechanism is not consistent between development, evaluation and deployment. Such situations are called "missingness shift", and may occur especially when measures unavailable at deployment are included in the development step to prevent missingness. The missingness mechanism at development is then no longer representative of the missingness mechanism at deployment. The distortion induced may be large enough to cancel the gain of information brought by the inclusion of the observation pattern, whereas a consistent MU prediction will be unaffected by the missingness shift \citep{smeden_cautionary_2020}. Deciding between MU or MC target when the risk of missingness shift is high requires therefore considering the trade-off between gain of information and risk of distortion induced by the shift. In any case, the target probability distribution for the forecaster should be clearly stated and determined before the model and the development procedure are designed.

\subsection{Relations between missingness mechanism assumptions}\label{sec:sub:relations-between-missingness-mechanism-assumptions}

Relations between the above-defined missingness mechanism assumptions are represented in figure \ref{fig:missingness-relations}. The demonstrations are available in Supplementary Material \ref{suppl:implication-relations}. Notably, while the MAR assumption implies MARX-YO, it does not imply MARX-YM, NIMO nor NICO. Also, both NIMO and MARX-YO are jointly necessary to achieve MARX-YM assumption. 

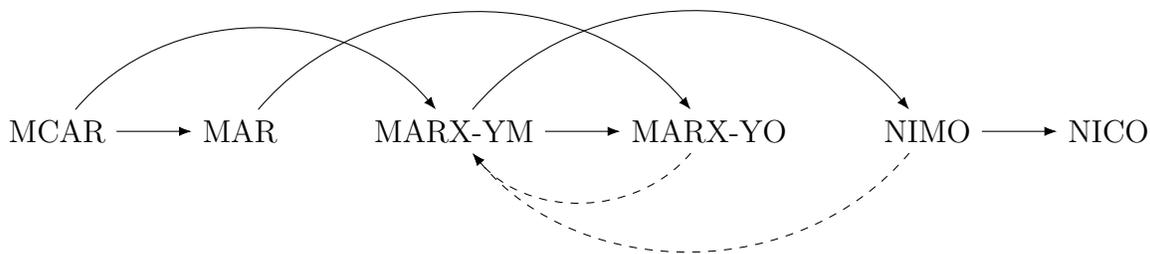
\begin{figure}[h]
    \centering
    \begin{tikzpicture}
    \node (MCAR) at (0,0) {MCAR};
    \node (MAR) [right = of MCAR] {MAR};
    \node (MARXYM) [right = of MAR] {MARX-YM};
    \node (MARXYO) [right = of MARXYM] {MARX-YO};
    \node (NIMO) [right = of MARXYO] {NIMO};
    \node (NICO) [right = of NIMO] {NICO};

    \path (MCAR) edge  (MAR);
    \path (MCAR) edge[bend left=50]  (MARXYM);
    \path (MAR) edge[bend left=50] (MARXYO);
    \path (MARXYM) edge (MARXYO);
    \path (MARXYM) edge[bend left=50] (NIMO);
    \path (NIMO) edge (NICO);
    \path[dashed] (NIMO) edge[bend left=50] (MARXYM);
    \path[dashed] (MARXYO) edge[bend left=50] (MARXYM);
\end{tikzpicture}
    \caption{Relations of implication between Missingness Completely at Random (MCAR), Missingness at Random (MAR), MAR for predictors - outcome missing (MARX-YM) and observed (MARX-YO), Non-Informative Missingness of the predictors for the outcome (NIMO) and Non-informative Complete Observation of the predictors for the outcome (NICO). Bold arrows represent direct implications. Dashed arrows represent joint implications.} 
    \label{fig:missingness-relations}
\end{figure}

\subsection{Illustrative scenarios}\label{sec:sub:illustrative-scenarios}

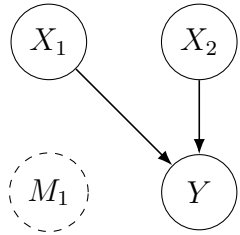
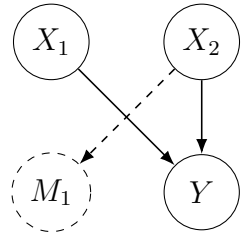
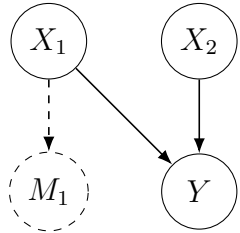
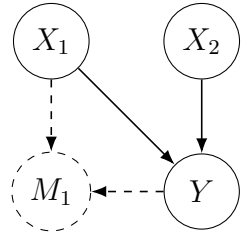
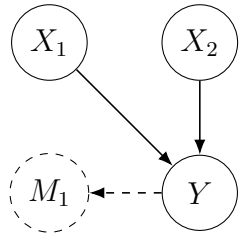
\begin{figure}[p]
\centering
\begin{subfigure}[t]{0.9\textwidth}
\footnotesize
\centering

\label{tab:properties}
\begin{threeparttable}
\begin{tabular}{cccccccc}
\hline
Scenario  & MCAR                  & MAR & MARX-YM                 & MARX-YO                 & NIMO                   & NICO & MC = MU? \\ \hline
1 & \cmark & \cmark & \cmark & \cmark & \cmark & \cmark & Yes                    \\
2 & \xmark & \cmark & \cmark & \cmark & \cmark & \cmark & Yes                    \\
3 & \xmark & \xmark & \xmark & \xmark & \xmark & \cmark & Yes\tnote{*}                  \\
4 & \xmark & \xmark & \xmark & \xmark & \xmark & \xmark & No                   \\
5 & \xmark & \xmark & \xmark & \cmark & \xmark & \xmark & No \\ \hline
\end{tabular}
\begin{tablenotes}
\footnotesize
\item[*] For complete cases only
\end{tablenotes}
\end{threeparttable}
\end{subfigure}
\par\bigskip
    \centering
    \begin{subfigure}[t]{0.4\textwidth}
        \centering
            \begin{tikzpicture}[node distance = 2cm,
state/.style={circle, draw, minimum size=1cm}]
        \node[state] (x1) {$X_1$};
        \node[state] (x2) [right of=x1] {$X_2$};
        \node[state] (y) [below of=x2] {$Y$};
        \node[state, dashed] (m1) [below of=x1] {$M_1$};
        
        \path[semithick] (x1) edge node {} (y);
        \path[semithick] (x2) edge node {} (y);
      \end{tikzpicture}
        \caption{Scenario 1}
    \end{subfigure}%
    ~ 
    \begin{subfigure}[t]{0.4\textwidth}
        \centering
        \begin{tikzpicture}[node distance = 2cm,
state/.style={circle, draw, minimum size=1cm}]
        \node[state] (x1) {$X_1$};
        \node[state] (x2) [right of=x1] {$X_2$};
        \node[state] (y) [below of=x2] {$Y$};
        \node[state, dashed] (m1) [below of=x1] {$M_1$};
        
        \path[semithick] (x1) edge node {} (y);
        \path[semithick] (x2) edge node {} (y);
        \path[semithick, dashed] (x2) edge node {} (m1);
      \end{tikzpicture}
        \caption{Scenario 2}
    \end{subfigure}

\par\bigskip

    \begin{subfigure}[t]{0.4\textwidth}
        \centering
            \begin{tikzpicture}[node distance = 2cm,
state/.style={circle, draw, minimum size=1cm}]
        \node[state] (x1) {$X_1$};
        \node[state] (x2) [right of=x1] {$X_2$};
        \node[state] (y) [below of=x2] {$Y$};
        \node[state, dashed] (m1) [below of=x1] {$M_1$};
        
        \path[semithick] (x1) edge node {} (y);
        \path[semithick] (x2) edge node {} (y);
        \path[semithick, dashed] (x1) edge node {} (m1);
      \end{tikzpicture}
        \caption{Scenario 3}
    \end{subfigure}%
    ~ 
    \begin{subfigure}[t]{0.4\textwidth}
        \centering
        \begin{tikzpicture}[node distance = 2cm,
state/.style={circle, draw, minimum size=1cm}]
        \node[state] (x1) {$X_1$};
        \node[state] (x2) [right of=x1] {$X_2$};
        \node[state] (y) [below of=x2] {$Y$};
        \node[state, dashed] (m1) [below of=x1] {$M_1$};
        
        \path[semithick] (x1) edge node {} (y);
        \path[semithick] (x2) edge node {} (y);
        \path[semithick, dashed] (x1) edge node {} (m1);
        \path[semithick, dashed] (y) edge node {} (m1);
      \end{tikzpicture}
        \caption{Scenario 4}
    \end{subfigure}

\par\bigskip

    \begin{subfigure}[t]{0.4\textwidth}
        \centering
            \begin{tikzpicture}[node distance = 2cm,
state/.style={circle, draw, minimum size=1cm}]
        \node[state] (x1) {$X_1$};
        \node[state] (x2) [right of=x1] {$X_2$};
        \node[state] (y) [below of=x2] {$Y$};
        \node[state, dashed] (m1) [below of=x1] {$M_1$};
        
        \path[semithick] (x1) edge node {} (y);
        \path[semithick] (x2) edge node {} (y);
        \path[semithick, dashed] (y) edge node {} (m1);
      \end{tikzpicture}
        \caption{Scenario 5}
    \end{subfigure}
\caption{\label{fig:illustrative-scenarios-dag}Directed Acyclic Graphs of five illustrative scenarios with two predictors and one outcome, and their missingness mechanism properties. Bold and dashed nodes represent model variables and missingness indicators, respectively.}
\end{figure}

Figure \ref{fig:illustrative-scenarios-dag} displays directed acyclic graphs under five scenarios of missingness mechanisms in the case of two predictors $X_1$ and $X_2$ and an outcome $Y$ with missing values on $X_1$ only. In all cases, $M_Y$ is assumed independent from both $Y$, $\X$ and $\Mx$.

The properties of each missingness mechanism with regards to MCAR, MARX-YM, MARX-YO, NIMO and NICO assumptions is detailed in the joint table.

\section{Procedures for handling missing values in prediction}\label{sec:methods}

The choice of a training procedure should account for whether missing predictors may occur at deployment, and which probability distribution is targeted.

\subsection{Training procedures for forecasters with missing predictors at deployment}\label{sec:sub:training-procedures-with-missing-predictors-at-deployment}

When missing predictors may be present at deployment, a forecaster defined only on fully observed predictors is insufficient. Methods must either (i) explicitly model the distribution of the missing predictors, or (ii) construct predictors defined directly on observed patterns. Existing approaches can be grouped into four  classes: pattern sub-models, marginalisation, imputation-based procedures, and missingness incorporated as attribute (MIA).

\subsubsection{Pattern Sub-models}\label{sec:subsub:pattern-sub-models}

Pattern Sub-models (PS) consist in fitting one forecaster per observation pattern of the predictors. Each sub-model estimates the MC probability distribution on the subset of the training data corresponding to that pattern. Under correct model specification, PS yields Bayes consistent MC prediction under arbitrary missingness mechanisms of the predictors, including MNAR, as demonstrated by \cite{fletcher_mercaldo_missing_2020}. 
Its main limitation is the number of sub-models to train, as the number of patterns may grow exponentially with the number of predictors, and rare patterns may lead to unstable estimation.

Complete Case Sub-models (CCS) \citep{fletcher_mercaldo_missing_2020} is an alternative to PS, training every sub-model on all the observations that have at least the  observed predictors of the pattern. This approach significantly increases the size of the training data subsets, and enables estimation of sub-models for unobserved patterns. CCS is not consistent for MC risk but has been shown consistent for MU risk only in restricted settings, for example when a single predictor is subject to missingness and NICO holds \citep{bartlett_asymptotically_2015}, but is not generally consistent (see Supplementary Material \ref{sec:ccs-is-inconsistent}).

\subsubsection{Marginalisation}\label{sec:marginalisation}

Marginalisation targets the MU probability distribution by integrating over the missing predictors:
\begin{equation*}
    \Pr(Y \mid \X_o) = \int \Pr(Y \mid \X_o, \X_m = \x_m)\Pr(\X_m = \x_m \mid \X_o)\, d\x_m
\end{equation*}

This approach requires estimating both the conditional outcome model $\Pr(Y \mid \X)$ and the conditional distribution of missing predictors $\Pr(\X_m \mid \X_o)$. Identification from observed data requires MAR, or MARX-YO when development is restricted to observations with observed outcomes (see section \ref{sec:subsub:Bayes-consistent-training-procedures-for-mu-risk}). Under these conditions and correct model specification, marginalisation of the missing predictors (either analytically or numerically using Monte Carlo approximation) yields Bayes consistent MU prediction. The addition of missingness indicators may lead to MC prediction, although this approach has not been studied to our knowledge.

\subsubsection{Imputation-Based Procedures}\label{sec:subsub:imputation}

Imputation replaces missing values by imputed values derived from observed variables. At development, imputation may be performed once (single imputation) or multiple times (Multiple Imputation, MI), sampling over the estimated distribution of the missing predictors given the observed predictors and the outcome. At development and validation, the imputation model may be estimated on available data, provided sufficient observations are present. At deployment, the imputation model must be provided alongside a forecaster based on complete predictors. An alternative named "stacked imputation" \citep{janssen_dealing_2009} includes at deployment the new individual in a larger dataset before performing MI on the whole set, but this procedure requires access to such a dataset and computational capacities.

When imputation models are correctly specified and MAR (or MARX-YO after exclusion of observations with missing outcome) holds, MI combined with standard estimation procedures yields Bayes consistent MU prediction. In this setting, MI can be interpreted as an empirical approximation of the marginalisation approach described above.

In contrast, deterministic imputation followed by regression (impute-then-regress procedures) has been shown to yield Bayes consistent MC prediction under arbitrary missingness mechanisms for almost all imputation functions, including unconditional imputation such as zero or unconditional mean, provided the class of forecasters is sufficiently rich \citep{le_morvan_whats_2021}. In this case, the procedure implicitly exploits the information carried by the observation pattern and therefore targets the MC probability distribution. Similarly, multiple imputation including missingness indicators (MIMI) has shown good performance in simulations \citep{hoogland_handling_2020, sperrin_multiple_2020, sisk_imputation_2023} and has been proved equivalent with PS \citep{fletcher_mercaldo_missing_2020}.

\subsubsection{Missingness Incorporated as Attribute (MIA)}\label{sec:subsub:mia}

Lastly, the "Missingness Incorporated as Attribute" (MIA) approach augments the domain of each predictor to include a specific value representing missingness \citep{twala_good_2008}. The prediction model is then estimated directly on these augmented predictors, which are fully observed by construction. MIA can be interpreted as a particular case of deterministic imputation mapping observations into an augmented predictor space. Under flexible modelling classes, it yields Bayes consistent MC prediction under arbitrary missingness mechanisms \citep{josse_consistency_2024}. Its practical performance ultimately depends, however, on whether the selected model class can adequately handle the resulting composite domains, especially for numeric predictors whose resulting domain is partly discrete.

\subsection{Training procedures when no missing predictors are expected at deployment}\label{sec:sub:training-procedures-no-missing-at-deployment}

When all predictors are expected to be observed at deployment, the prediction target corresponds to complete cases. In this setting, the distinction between MU and MC depends on whether NICO holds.

At development, MLE or MI are Bayes consistent MU training procedures under MAR, and under weaker conditions such as MARX-YO when estimation is restricted to observed outcomes. Complete Cases Analysis, MIA, MIMI or deterministic imputation yield consistent MC prediction under any missingness mechanism of the predictors. If NICO holds, MU and MC coincide on complete cases and these methods are therefore also consistent for MU prediction. If neither MARX-YO nor NICO hold, consistent MU prediction is generally not achievable without explicit modelling of the missingness mechanism.

\section{Simulation study}\label{sec:simulations}

In the previous section, we characterised several methods for handling missing predictors and identified the assumptions under which they yield Bayes consistent MU or MC prediction. In this section, we further illustrate these results by comparing their performance on simulated data with a predictor potentially missing under various missingness mechanisms.

\subsection{Methods}\label{sec:sub:simulation-methods}

\subsubsection{Data generation}\label{sec:subsub:simulation-data-generating}

We studied the five scenarios of missingness mechanisms described in figure \ref{fig:illustrative-scenarios-dag} with two continuous predictors $X_1, X_2$ and a continuous outcome $Y$. The joint distribution $\Pr(Y, X_1, X_2)$ was kept identical across scenarios, and the conditional distribution $\Pr(M_1 \mid Y, X_1, X_2)$ was adapted with respect to the missingness scenario and the target proportion of missing values. For all scenarios, $M_Y$ was assumed independent from $Y, \X, \Mx$ and always observed at development. For each missingness mechanism and each missingness proportion, independent training and testing datasets of 1,000 observations each were generated. Expected missingness proportion was increased from 0 to 70\%, by increments of 0.1\%, resulting overall in 701 couples of datasets for each scenario.

\subsubsection{Procedures evaluated}\label{sec:subsub:simulation-procedures-evaluated}

Pattern sub-models (PS), complete-cases sub-models (CCS), Multiple Imputation without (MI) or with (MIMI) missingness indicators, and MLE with marginalisation without (MLE-M) or with (MLEMI-M) missingness indicators were assessed, using linear models without interaction terms. Marginalisation was performed using Monte Carlo approximation.

According to the framework developed in Sections 2–3, we expected: (i) divergence between MU and MC Bayes optimal predictions when NIMO did not hold, (ii) MI and MLE-M to be consistent for MU under MAR and MARX-YO, (iii) PS, MIMI, and MLEMI-M to be consistent for MC under all mechanisms.

Missingness Incorporated as Attribute (MIA) was not assessed, as the MIA approach with continuous variables requires specific approaches such as tree-based methods. Constant and unconditional mean imputations were not assessed because these procedures distort the distribution of the predictors, causing failure of linear models. Performance of such methods using tree-based methods have already been assessed by other authors for both MIA and constant imputation \citep{josse_consistency_2024, gao_comparing_2025}.

\subsubsection{Estimands and performance measures}\label{sec:subsub:simulations-estimands}

The estimand under study was the outcome value on the test set. For all scenarios, and all missingness proportions, each training procedure was applied on the training set, resulting in a forecaster which was deployed on the testing set. The outcome predictions obtained on the test set were then compared to the true generated outcome using MSE.

Results are reported as scatter plots of MSE according to the missingness proportion over the 1000 replicates, and a smoothing curve per procedure (Locally Estimated Scatterplot Smoothing (LOESS) with span parameter of 0.5) is added  to improve readability. The MU and MC probabilities are also plotted as reference to represent the performance achievable by a Bayes optimal predictor.

\subsubsection{Subgroup and supplementary analyses}\label{sec:subsub:simulation-methods-exploratory-analysis}

Subgroup analyses were performed for complete and incomplete subgroups separately. To assess the necessity of excluding the observations with missing $Y$ when estimating MU prediction models in MNAR and MARX-YO scenarios, an exploratory analysis was conducted by introducing missingness of $Y$ (completely independent from the values of $Y$, $X_1$, $X_2$ and $M_{X_1}$) in the training set of scenario 5. Performance of MI and MLE-M were compared with and without exclusion of observations with missing $Y$ from the training set.

\subsubsection{Software}\label{sec:subsub:simulation-methods-software}

Analyses were conducted with R language (version 4.5.0), using \textit{mice} and \textit{norm} packages \citep{R-4.5.0_2025, vanBuuren_mice_2011, schafer_norm_2024}. The code used for the simulation and figures, and the obtained datasets are available in the dedicated repository \citep{catoire_zenodo_2026}.

\subsection{Results}\label{sec:sub:simulation-results}

\subsubsection{Main analysis}\label{sec:subsub:simulation-main}

\begin{figure}[p]
  \centering

  \includegraphics[width=\linewidth]{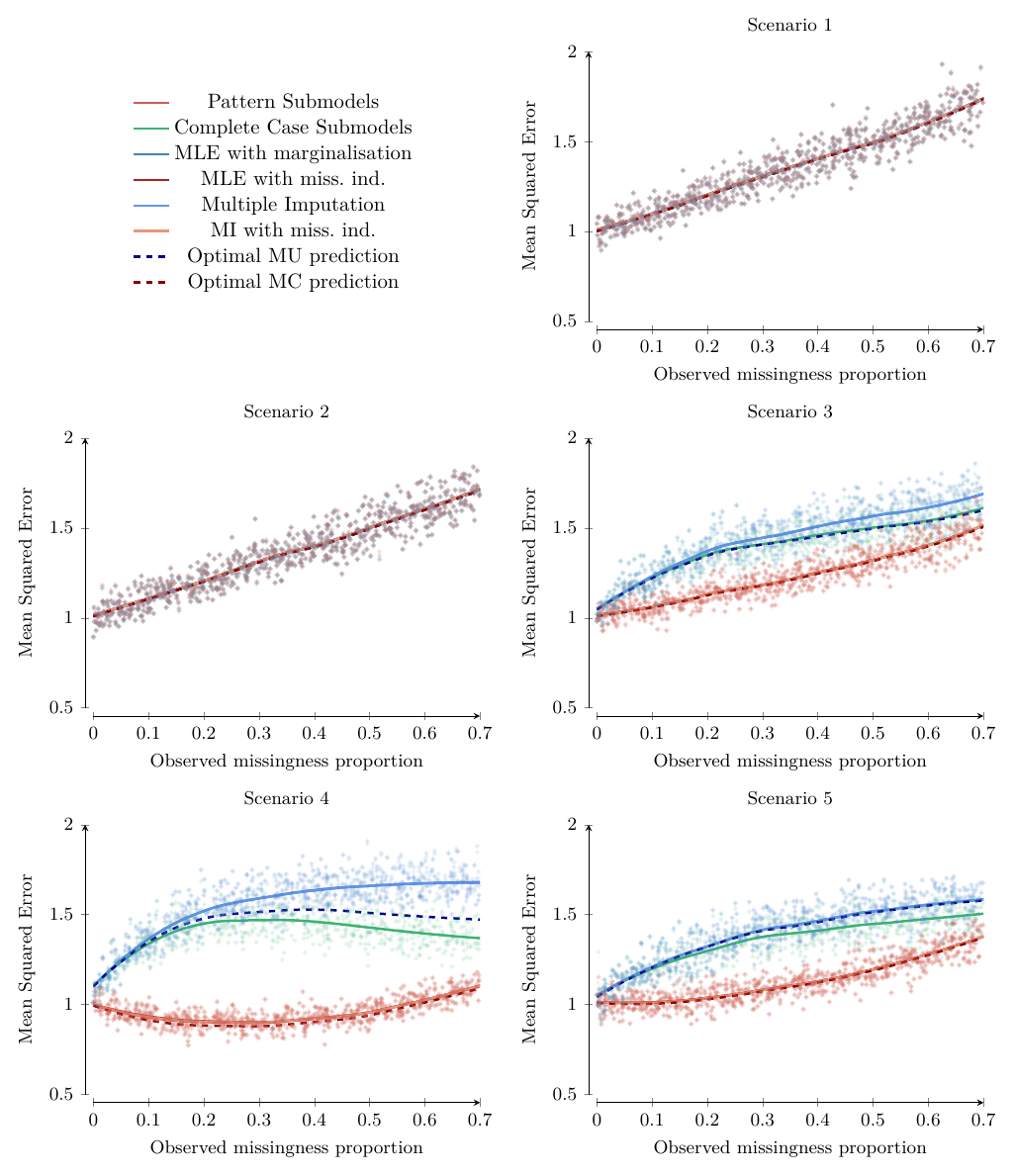}
  \caption{\label{fig:all_scenarios_continuous}Mean Squared Error (MSE) for various proportions of missing values of evaluated procedures for the five scenarios, for continuous variables}
\end{figure}

Figure \ref{fig:all_scenarios_continuous} shows the MSE of each procedure for the five considered scenarios. The MU and MC Bayes optimal prediction obtained identical performances for scenarios 1 and 2, and diverge for scenarios 3 to 5, as expected with the NIMO assumption. When MU and MC diverged (scenarios 3–5), the MC-Bayes optimal forecaster achieved lower risk, reflecting the additional information carried by the observation pattern.

PS, MIMI and MLEMI-M showed results very close to the MC-Bayes optimal predictor, across all scenarios including MNAR. MSE of Complete case sub-models (CCS) were close to the optimal prediction for scenarios 1 and 2. For scenario 3, CCS performance were close to the MU optimal predictor. For scenarios 4 and 5, CCS results diverged from both MU and MC optimal predictors.

MI and MLE-M showed results very close to the Bayes optimal predictors for scenarios 1 and 2, as expected under MAR assumption. They diverged from both MU and MC predictions for scenarios 3 and 4. For scenario 5 (MNAR but MARX-YO), they were very close to the MU-Bayes optimal predictor, as expected with the consistency under MARX-YO (section \ref{sec:subsub:Bayes-consistent-training-procedures-for-mu-risk}).

\subsubsection{Subgroup analysis}\label{sec:subsub:simulation-subgroup-results}

Results of subgroup analyses for complete and incomplete observations are described in Supplementary Material \ref{suppl:secondary-analyses} (Figures \ref{fig:all_scenarios_continuous_mse_complete} and \ref{fig:all_scenarios_continuous_mse_incomplete}). For scenarios 1 and 2, MU and MC Bayes optimal predictors were equal for both subgroups. For scenario 3, they diverged only for the incomplete cases, consistent with the NICO assumption. They diverged in both subgroups for scenarios 4 and 5.

PS, MIMI and MLEMI-M showed MSEs very close to MC-Bayes optimal predictors across subgroups for all missingness mechanisms. MI and MLE-M performances were close to MU-Bayes optimal predictors across all scenarios for incomplete cases. For complete cases, they diverged from MU-Bayes optimal prediction in scenarios 3 and 4. CCS performances were close to Bayes optimal prediction in scenarios 1 and 2 (MAR). In scenario 3, they were close to MU-Bayes optimal prediction in the complete subgroup, and close to the MU-Bayes optimal prediction in the incomplete subgroup. In scenario 4 and 5, CCS showed performances close to MC-Bayes optimal prediction in the complete subgroup and close to MU-Bayes optimal prediction in the incomplete subgroup.

\subsubsection{Supplementary analysis under MNAR in Scenario 5}\label{sec:subsub:simulation-explo-results}

In the MNAR, MARX-YO scenario 5, we assessed the imputation technique according to the development data used (figure \ref{fig:results-explo}). When estimated on the whole dataset including observations with missing $Y$, the forecaster diverged from the MU-Bayes optimal prediction. In contrast, when estimated on the subset of observed $Y$, the forecaster performed virtually the same as the MU-Bayes optimal prediction.

\section{Application to emergency medicine}\label{sec:application}

As an empirical illustration, we applied the previously described procedures to predict the risk of severe traumatic injury after a high-energy accident using clinical predictors collected at patient admission in the emergency department. Prediction in this context may support triage and early decision-making. However, several predictors, including potentially informative ones, may be unavailable. For example, a patient may be unable to report a coagulation disorder because they are unconscious. A model applicable only when all predictors are observed would therefore either exclude clinically relevant variables or be restricted to a limited subset of patients. Methods robust to missing predictors at deployment may therefore be preferable. In addition, some predictors may not be measured when the patient initially appears stable. For example, pulse oximetry may be omitted in patients who do not appear severely injured. In such cases, missingness itself may carry information in the development dataset. However, if measurement of a given predictor becomes routine in future practice, its observation pattern may change, illustrating the possibility of missingness shift.

\subsection{Data from the Bordeaux Hospital}\label{sec:sub:application-setup}

The data were retrospectively collected from patients admitted to a level-one trauma centre in Bordeaux, France, following a high-velocity trauma requiring trauma bay admission and whole-body computed tomography between 2017 and 2019. Patients under 18 years of age were excluded. Data were extracted from the electronic medical records, based on entries recorded by physicians and nurses during initial management. The total number of patients was 678. Data collection followed established regulatory guidelines for the protection of human subjects.

The outcome was severe trauma, defined as an Injury Severity Score (ISS) of 15 or more. The ISS was determined after full injury assessment, including imaging. The predictors considered were age (always observed), altered mental status (Glasgow Coma Scale < 15), coagulation disorder, and hypoxaemia (defined as requirement of oxygen to maintain pulse oximetry > 96\%).

\subsection{Training procedures and evaluation}\label{sec:sub:application-training}

We evaluated Pattern Sub-models (PS), Multiple Imputation without missingness indicators (MI), and Multiple Imputation with missingness indicators (MIMI). All procedures used logistic regression without interaction terms. Predictive performance was assessed using leave-one-out cross-validation. For each individual, the model was trained on the remaining observations and used to generate a predicted value for that individual. Performance was measured using the Brier score, with 95\% confidence intervals obtained by non-parametric bootstrap with 10,000 replicates. Results are reported overall and stratified by observation pattern.

Analyses were conducted using R (version 4.5.0 \cite{R-4.5.0_2025}) with the \textit{mice} and \textit{naniar} packages \citep{vanBuuren_mice_2011, tierney_naniar_2021}. The analysis code is available in a dedicated repository \citep{catoire_zenodo_2026}. The dataset contains individual health data and is therefore not publicly available. Access may be granted upon reasonable request and with appropriate ethical and institutional approvals.

\subsection{Results}\label{sec:sub:application-results}

\begin{figure}[p]
    \centering
    \includegraphics[width=\linewidth]{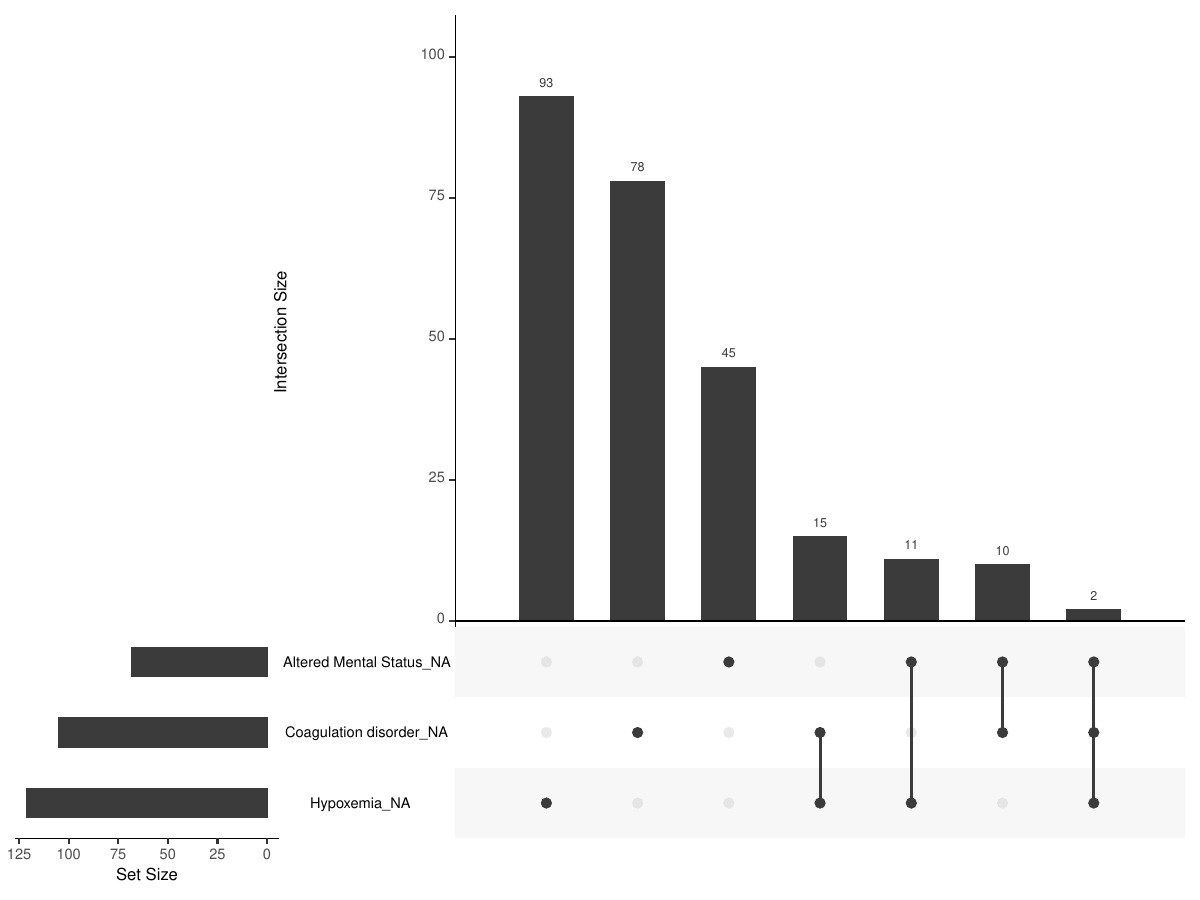}
    \caption{Distribution of the observation patterns. From \textit{naniar} package \cite{tierney_naniar_2021}.}
    \label{fig:missingness-patterns}
\end{figure}

Among the 678 patients included, 147 (21.7\%) had severe trauma (see description in Table \ref{tab:application-population-characteristics}). Patients with severe trauma were older and more frequently presented altered mental status and coagulation disorder. Missingness of altered mental status and hypoxaemia was more frequent among patients with severe trauma. The distribution of observation patterns is given in Figure \ref{fig:missingness-patterns}. Overall, 424 patients (62.5\%) had all predictors observed. Among the 254 patients (37.5\%) with at least one missing predictor, 216 (31.9\%) had one predictor missing, 36 (5.3\%) had two missing, and 2 (0.3\%) had three missing.

Brier scores for each procedure are reported in Table  \ref{tab:application-results}, overall and stratified by observation pattern. The overall Brier scores were 0.165 [0.147–0.183] for PS, 0.228 [0.218–0.239] for MI, and 0.299 [0.286–0.311] for MIMI. The best performance of Pattern Sub-models suggest that the observation pattern carries predictive information. However, these performances were mostly driven by its superiority for the completely observed pattern. Its performance was lower than MI for the rare patterns likely reflecting smaller effective training sample sizes. MIMI performed worse than MI for almost all the patterns.

\begin{table}[p]
\caption{Population characteristics (N=678) \label{tab:application-population-characteristics}
}
\centering
\small
\centering
\begin{tabular}[t]{cccc}
\hline
  & No trauma & Trauma & Overall\\
 & (N=531) & (N=147) & (N=678)\\
 \hline
Age, median [IQR] & 37.0 [25.0-51.0] & 44.0 [29.0-57.0] & 39.0 [26.0-52.0]\\
Altered mental status & 40 (7.5\%) & 26 (17.7\%) & 66 (9.7\%)\\
Missing & 45 (8.5\%) & 23 (15.6\%) & 68 (10.0\%)\\
Hypoxemia & 41 (7.7\%) & 9 (6.1\%) & 50 (7.4\%)\\
Missing & 73 (13.7\%) & 48 (32.7\%) & 121 (17.8\%)\\
Coagulation disorder & 36 (6.8\%) & 19 (12.9\%) & 55 (8.1\%)\\
Missing & 80 (15.1\%) & 25 (17.0\%) & 105 (15.5\%)\\
\hline
\end{tabular}
\end{table}
\clearpage

\clearpage

\begin{table}[p]
    \centering
\caption{Brier scores (95\%CI) of Pattern Sub-models (PS), multiple imputation (MI), multiple imputation with missingness indicators (MIMI) overall and among subgroups per observation pattern. \label{tab:application-results}}
\footnotesize
\centering
\begin{threeparttable}
\begin{tabular}[t]{ccccccccc}
\hline
\multicolumn{4}{c}{Pattern}  & \multirow{2}{*}{N (\%)} & \multirow{2}{*}{PS}  & \multirow{2}{*}{MI}  & \multirow{2}{*}{MIMI} \\
Age & AMS\tnote{a} & Hypox.\tnote{b} & Coag.\tnote{c} & & & & \\
\hline
\multicolumn{4}{c}{Overall} & 678 (100.0\%) & 0.165 (0.147–0.183) & 0.228 (0.218–0.239) & 0.299 (0.286–0.311)\\
\cmark & \cmark & \cmark & \cmark & 424 (62.5\%) & 0.129 (0.106–0.152) & 0.229 (0.218–0.241) & 0.298 (0.285–0.312)\\
\cmark & \cmark & \xmark & \cmark & 93 (13.7\%) & 0.227 (0.186–0.269) & 0.242 (0.221–0.264) & 0.242 (0.195–0.290)\\
\cmark & \cmark & \cmark & \xmark & 78 (11.5\%) & 0.178 (0.127–0.231) & 0.207 (0.177–0.238) & 0.345 (0.308–0.386)\\
\cmark & \xmark & \cmark & \cmark & 45 (6.6\%) & 0.252 (0.199–0.307) & 0.240 (0.170–0.314) & 0.340 (0.280–0.399)\\
\cmark & \cmark & \xmark & \xmark & 15 (2.2\%) & 0.343 (0.232–0.476) & 0.236 (0.188–0.288) & 0.238 (0.146–0.336)\\
\cmark & \xmark & \xmark & \cmark & 11 (1.6\%) & 0.311 (0.185–0.452) & 0.263 (0.138–0.394) & 0.227 (0.155–0.301)\\
\cmark & \xmark & \cmark & \xmark & 10 (1.5\%) & 0.117 (0.012–0.317) & 0.101 (0.026–0.241) & 0.469 (0.376–0.538)\\
\cmark & \xmark & \xmark & \xmark & 2 (0.3\%) & 0.500 (0.000–1.000) & 0.325 (0.081–0.569) & 0.291 (0.262–0.319)\\
\hline
\end{tabular}
\begin{tablenotes}
\footnotesize
\item[a] AMS: Altered Mental Status
\item[b] Hypox.: hypoxemia
\item[c] Coag.: coagulation disorder
\item Note: Confidence interval estimation of rare patterns using bootstrapping may not be reliable, and are indicative.
\end{tablenotes}
\end{threeparttable}
\end{table}
\clearpage

\section{Discussion and conclusion}\label{sec:discussion}

This works contributes to clarifying the role of missingness in prediction by distinguishing two prediction targets: the "Missingness-Unconditional" (MU) probability distribution $\Pr(Y \mid \X_o)$, and the "Missingness-Conditional" (MC) probability distribution $\Pr(Y \mid \X_o, \Mx)$. This distinction offers an explanation to apparent contradictions in the literature, where training procedures considered biased for inference may nevertheless achieve optimal predictive performance. The key insight of this work is that these procedures may target a different estimand.

We introduced a nomenclature of missingness mechanisms tailored to prediction, including NIMO, NICO, MARX-YM and MARX-YO assumptions. NIMO ensures equivalence between MU and MC probabilities for all observation patterns, and NICO ensures equivalence if all predictors are observed. A central result is that consistent MC prediction can be achieved under arbitrary missingness mechanism of the predictors including MNAR, using procedures such as PS, MIA, deterministic imputation, MIMI or MLE with missingness indicators. Opposite to this, consistent MU prediction using MLE or MI without missingness indicators generally requires MAR, or MARX-YO when training data is restricted to cases with observed outcome.

The choice between MU and MC target should be driven by the missingness mechanism assumed and the risk of inconsistency of the missingness mechanism between development and deployment (missingness shift). Prediction model design should also consider the possibility of missing predictors at deployment, which requires dedicated handling methods.

Simulation results were fully consistent with the theoretical framework. MU and MC probability distributions coincided under NIMO, and diverged otherwise. When they diverged, MC-Bayes optimal forecaster achieved lower risk, reflecting the additional information carried by the observation pattern. Methods conditioning on the observation pattern achieved performance close to the MC-Bayes optimal forecaster under all mechanisms. In contrast, methods targeting MU probability distribution were close from the MU-Bayes optimal predictor only under MARX-YO and diverged otherwise. Complete Case Sub-models showed acceptable performance only in restricted settings.

In the application on severe trauma prediction, Pattern Sub-models achieved the lowest overall Brier score, suggesting that the observation pattern carried predictive information. However, PS performed less well in rare patterns, likely due to smaller sample sizes. MIMI performed worse than both PS and MI, in contrast with simulations where PS and MIMI were equivalent. This discrepancy likely reflects misspecification of the model, particularly from the absence of interaction terms between predictors and missingness indicators.

Our results are consistent with prior theoretical and empirical findings. \cite{fletcher_mercaldo_missing_2020} showed that PS achieve optimal prediction under arbitrary missingness mechanisms, and \cite{le_morvan_whats_2021} obtained similar results for deterministic imputation. Our framework clarifies that these procedures target MC rather than MU, resolving the apparent contraction with classical inference theory where MAR is required for unbiased parameter estimation. Similarly, empirical findings on missing indicators \citep{hoogland_handling_2020, sisk_imputation_2023} are consistent with our interpretation, as these approaches exploit missingness information and therefore target MC.

Several limitations must be acknowledged. First, simulations were conducted in simple settings with a single potentially missing predictor and linear models without interactions. More complex settings, with more variables or mixed predictor types, or use of flexible learners (e.g. tree-based or neural networks) may introduce additional nuances, although the theoretical results are not dependent from type of predictor or class of learner. Calibration and effect of missingness shift were not investigated.

In conclusion, missingness in prediction fundamentally differs from inference as observation pattern may carry relevant information on the outcome. The proposed paradigm provides a structured framework for selecting appropriate methods and evaluate the impact of missingness at development, validation and deployment, from which recommendations are suggested in table \ref{tab:recommandations}. Further methodological development may include diagnostic tests for missingness mechanism assumptions, the formal evaluation of calibration under missingness predictors, the effect of missingness shift on prediction performance, and the impact of the present results on the validation strategy for predictive algorithms.

\section{Disclosure statement}\label{sec:disclosure-statement}

The authors declare that there are no conflicts of interest regarding the publication of this article.

\section{Data Availability Statement}\label{sec:data-availability-statement}
The code used for the simulation study and real-data application, along with the simulated datasets, are provided in the Supplementary Material. The real-world clinical dataset used in the application section cannot be publicly shared due to healthcare data protection regulations. Access may be granted upon reasonable request and subject to compliance with applicable ethical approvals and data access regulations.

\section{Acknowledgments}

This study was carried out within the framework of the University of Bordeaux’s France 2030 program - Impulsion Public Health Data Science Research Network (RRI PHDS).

\begin{table}[p]
\centering
\caption{\label{tab:recommandations}Recommended methods for handling missing predictors}
\begin{threeparttable}
\small
\setlength{\tabcolsep}{5pt}

\definecolor{headerdark}{HTML}{7F7F7F}
\definecolor{headerlight}{HTML}{C9C9C9}
\definecolor{rowgray}{HTML}{F2F2F2}

\begin{tabular}{ccccl}

\rowcolor{headerdark}
\multicolumn{5}{c}{\color{white}\textbf{Models forbidding missing predictors at deployment}}\\
\rowcolor{headerlight}
\textbf{MAR} & \textbf{MARX-YO} & \textbf{NICO} & \textbf{Target} & \textbf{Available methods} \\
\cmark & \cmark & \cmark & MU = MC 
& MLE\tnote{a}, MI\tnote{a}, CCA, MIA, DI, MIMI \\

\rowcolor{rowgray}
\cmark & \cmark & \xmark & MU 
& MLE\tnote{a}, MI\tnote{a} \\

\cmark & \cmark & \xmark & MC 
& CCA, MIA, DI, MIMI \\

\rowcolor{rowgray}
\xmark & \cmark & \xmark & MU 
& MLE\tnote{b}, MI\tnote{b} \\

\xmark & \cmark & \xmark & MC 
& CCA, MIA, DI, MIMI \\

\rowcolor{rowgray}
\xmark & \xmark & \cmark & MU = MC 
& CCA, MIA, DI, MIMI \\

\xmark & \xmark & \xmark & MU 
& --- \\

\rowcolor{rowgray}
\xmark & \xmark & \xmark & MC 
& CCA, MIA, DI, MIMI \\
\rowcolor{headerdark}
\multicolumn{5}{c}{\color{white}\textbf{Models allowing missing predictors at deployment}}\\
\rowcolor{headerlight}
\textbf{MAR} & \textbf{MARX-YO} & \textbf{NIMO} & \textbf{Target} & \textbf{Available methods} \\

\cmark & \cmark & \cmark & MU = MC 
& MLE\tnote{a} (marg), MI\tnote{a} (imp), PS, MIA, DI, MIMI \\

\rowcolor{rowgray}
\cmark & \cmark & \xmark & MU 
& MLE\tnote{a} (marg), MI\tnote{a} (imp) \\

\cmark & \cmark & \xmark & MC 
& PS, MIA, DI, MIMI \\

\rowcolor{rowgray}
\xmark & \cmark & \xmark & MU 
& MLE\tnote{b} (marg), MI\tnote{b} (imp) \\

\xmark & \cmark & \xmark & MC 
& PS, MIA, DI, MIMI \\

\rowcolor{rowgray}
\xmark & \xmark & \cmark & MU = MC 
& PS, MIA, DI, MIMI \\

\xmark & \xmark & \xmark & MU 
& --- \\

\rowcolor{rowgray}
\xmark & \xmark & \xmark & MC 
& PS, MIA, DI, MIMI \\
\end{tabular}

\footnotesize
\begin{tablenotes}
\tiny
\item MU: Missingness-Unconditional probability. MC: Missingness-Conditional probability. MLE: Maximum Likelihood Estimation. MI: Multiple Imputation. CCA: Complete Cases Analysis. MIA: Missingness Incorporated as Attribute. DI: Deterministic Imputation. MIMI: Multiple Imputation with Missingness Indicators. PS: Pattern Submodels.
\item marg. Requires marginalisation over missing predictors at deployment.
\item imp. Requires imputation of missing predictors at deployment.
\item[a] Uses all observations at development, including those with missing $Y$ (valid under MAR).
\item[b] Restricts development to observations with observed $Y$ (required under MARX-YO).
\end{tablenotes}

\end{threeparttable}
\end{table}

\clearpage
\bibliography{bibliography.bib}

\phantomsection
\bigskip

\begin{center}

\clearpage
{\large\bf SUPPLEMENTARY MATERIAL}

\end{center}

\begin{description}
\item[\ref{sec:sub:consistency-MLE-MARXYO}. Consistency of MLE under MARX-YO:]
Demonstrations of consistency of maximum-likelihood estimation procedure under MARX-YO.
\item[\ref{suppl:implication-relations}. Implication relations:] Demonstrations of implication relations between missingness mechanism assumptions.
\item[\ref{sec:ccs-is-inconsistent}. Consistency conditions of Complete Case Sub-models:] Demonstration of general inconsistency of Complete Cases Analysis under non-NICO mechanism or multiple missing predictors.
\item[\ref{suppl:secondary-analyses}. Secondary and exploratory analyses:] Secondary and exploratory analyses of the simulation study
\end{description}

\clearpage
\appendix

\section{Consistency of MLE on observations with observed outcome under MARX-YO}
\label{sec:sub:consistency-MLE-MARXYO}

\textbf{Proposition.} Assume that the following contiditional independences hold:

\begin{enumerate}
    \item $\Mx \perp \X_m \mid \X_o, Y$ \hfill (MARX-YO),
    \item $\Mx \perp M_Y \mid \X_o, \X_m, Y$,
    \item $M_Y \perp Y \mid \X_o$.
\end{enumerate}

Assume further that there is a model parameterised by $\boldsymbol{\eta}$
\begin{equation*}
\Pr_{\boldeta}(Y,\X \mid M_Y = 0)
\end{equation*}

that is correctly specified and identifiable, with true parameter values $\boldeta_0$ such that
\begin{equation*}
\Pr_{\boldeta_0}(Y,\X \mid M_Y = 0)
=
\Pr(Y,\X \mid M_Y = 0),
\end{equation*}
and that standard regularity conditions for consistency of maximum likelihood estimators hold. Assume moreover that the vector of parameters $\boldeta$ indexing $\Pr_{\boldeta}(Y,\X \mid M_Y=0)$ is distinct from the vector of parameters $\phi$ indexing $\Pr_{\phi}(\Mx \mid Y,\X,M_Y=0)$.

Then the training procedure $\Tr$ based on maximum likelihood estimation applied to the subsample
\begin{equation*}
\Data^{N_0}_{M_Y=0}
=
\{(Y_i,\X_{o,i},\M_{\X,i}) : M_{Y,i}=0\}
\end{equation*}
is MU–Bayes consistent:
\begin{equation*}
\Risk\left(\Tr(\Data^{N_0}_{M_Y=0})\right)
\xrightarrow{\textrm{P}}
\Risk^*_{\mathrm{MU}}
\qquad \text{as } N \to \infty
\end{equation*}

\textbf{Proof.}

From MARX-YO and $\Mx \perp M_Y \mid \X_o,\X_m,Y$, contraction \citep{dawid_conditional_1979} gives
\begin{equation*}
\Mx \perp (\X_m, M_Y) \mid \X_o, Y,
\end{equation*}
and weak union yields
\begin{equation*}
\Mx \perp \X_m \mid \X_o, Y, M_Y.
\tag{A1}
\end{equation*}

Consider the factorisation of the full-data distribution conditionally on $M_Y=0$:
\begin{align*}
\Pr(Y,\X_o,\X_m,\Mx \mid M_Y=0) &= \Pr(Y,\X_o,\X_m \mid M_Y=0)\Pr(\Mx \mid Y,\X_o,\X_m,M_Y=0) \\
&= \Pr(Y,\X_o,\X_m \mid M_Y=0)\Pr(\Mx \mid Y,\X_o,M_Y=0) \quad \textrm{(from A1)}
\end{align*}

Integrating out $\X_m$ yields the observed-data distribution
\begin{align*}
\Pr(Y,\X_o,\Mx \mid M_Y=0)
&=
\int
\Pr(Y,\X_o,\X_m = \x_m,\Mx \mid M_Y=0)\, d\x_m \\
&= \int \Pr(Y,\X_o,\X_m= \x_m \mid M_Y=0)\Pr(\Mx \mid Y,\X_o,M_Y=0) \ d\x_m \\
&=  \Pr(Y,\X_o \mid M_Y=0) \Pr(\Mx \mid Y,\X_o,M_Y=0)
\end{align*}

By denoting $\Pr_{\boldsymbol{\phi}}(\Mx \mid Y,\X,M_Y=0)$ a well-specified model for observation pattern parameterized by ${\boldsymbol{\phi}}$, the observed likelihood based on $\Data^{N_0}_{M_Y=0}$ factorises as
\begin{equation*}
\prod_{i:M_{Y,i}=0}
\Pr_{\boldeta}(Y_i,\X_{o,i} \mid M_Y = 0)
\;
\Pr_{\phi}(\M_{\X,i} \mid Y_i,\X_{o,i},M_Y=0),
\end{equation*}
Because the parameters in the two components of the likelihood are distinct, maximisation with respect to $\boldeta$ is unaffected by the missingness component; the predictor missingness is therefore ignorable for estimation of $\boldeta$.

Under standard MLE consistency results \citep{vandervaart_asymptotic_1998}, for $\mathcal{I}_{\textrm{MLE}}$ an inference procedure estimating $\boldeta$ using MLE on the subset of data with observed outcome $\Data^{N_0}_{M_Y=0}$, we have
\begin{equation*}
\mathcal{I}_{\textrm{MLE}}(\Data^{N_0}_{M_Y=0}) = \hat{\boldeta}^{N_0}
\xrightarrow{\textrm{P}}
\boldeta,
\tag{A2}
\end{equation*}

Consider $\Tr_{\textrm{MLE}}$ the training procedure yielding a forecaster $\hat\pr_{\textrm{MLE}}$ such that:
\begin{equation*}
    \hat\pr^{N_0}_{\textrm{MLE}}(\x_o) =
    \frac
    {\Pr_{\hat{\boldeta}}(Y, \X_o = \x_o \mid M_Y = 0)}
    {\int Pr_{\hat{\boldeta}}(Y = y, \X_o = \x_o \mid M_Y = 0) \ dy}
\end{equation*}

As the estimated parameter $\hat\boldeta$ converges in probability to $\boldeta$ (A2), the continuous mapping theorem yields:
\begin{equation*}
\Risk\left(
\hat\pr^{N_0}_{\textrm{MLE}}
\right)
\xrightarrow{\textrm{P}}
\Risk\left(\Pr(Y \mid \X_o, M_Y = 0)\right)
\qquad \text{as } N \to \infty
\end{equation*}

Provided also that
\begin{equation*}
M_Y \perp Y \mid \X_o
\quad \Rightarrow \quad
\Pr(Y \mid \X_o, M_Y=0)
=
\Pr(Y \mid \X_o),
\end{equation*}

We obtain:
\begin{equation*}
\Risk\left(
\hat\pr^{N_0}_{\textrm{MLE}}
\right)
\xrightarrow{\textrm{P}}
\Risk^*_{\mathrm{MU}}
\qquad \text{as } N \to \infty
\end{equation*}

\hfill $\square$

\section{Implication relations between missingness mechanism properties}
\label{suppl:implication-relations}

All implications below rely on the standard semi-graphoid properties of conditional independence: symmetry, decomposition, weak union, and contraction \citep{dawid_conditional_1979}. Although these properties do not hold for arbitrary context-specific conditional independencies, the statements of the form $\Mx = \mx \perp B \mid C$ can be interpreted as independence between the indicator variable $\mathbb{I}\{\Mx = \mx\}$ and $B$ conditionally on $C$, which implies valid properties of conditional independence. Strict positivity of conditioning sets is assumed.

\subsection{MCAR implies MARX-YM}

Under MCAR,
\begin{equation*}
(\Mx, M_Y) \perp (\X, Y)
\end{equation*}

By decomposition,
\begin{equation*}
\Mx \perp (\X, Y)
\end{equation*}

Partitioning $\X = (\X_o, \X_m)$ and applying weak union yields
\begin{equation*}
\Mx \perp (\X_m, Y) \mid \X_o
\end{equation*}
which is the definition of MARX-YM.

\subsection{MAR implies MARX-YO}

Under MAR, when $M_Y = 1$,
\begin{equation*}
    \Mx=\mx, M_Y = 1 \perp \X_m(\mx), Y \mid \X_o(\mx)
\end{equation*}

By weak union,
\begin{equation*}
    \Mx=\mx, M_Y = 1 \perp \X_m(\mx) \mid \X_o(\mx), Y
\end{equation*}

Which also holds when $M_Y = 0$ from MAR definition. As $M_Y$ is binary, this property holds for all values of $M_Y$:
\begin{equation*}
    \Mx=\mx, M_Y \perp \X_m(\mx) \mid \X_o(\mx), Y
\end{equation*}

By decomposition,
\begin{equation*}
    \Mx=\mx \perp \X_m(\mx) \mid \X_o(\mx), Y
\end{equation*}

Which is the definition of MARX-YO.

\subsection{MARX-YM implies MARX-YO}

Under MARX-YM,
\begin{equation*}
\Mx=\mx \perp \X_m(\mx), Y \mid \X_o(\mx)
\end{equation*}

By weak union,
\begin{equation*}
\Mx=\mx \perp \X_m(\mx) \mid \X_o(\mx), Y
\end{equation*}

which is MARX-YO.

\subsection{MARX-YM implies NIMO}

From MARX-YM,
\begin{equation*}
\Mx=\mx \perp \X_m(\mx), Y \mid \X_o(\mx)
\end{equation*}

By decomposition,
\begin{equation*}
\Mx=\mx \perp Y \mid \X_o(\mx)
\end{equation*}

which is NIMO.

\subsection{NIMO implies NICO}

Under NIMO,
\begin{equation*}
Y \perp \Mx=\mx \mid \X_o(\mx)
\quad \text{for all } \mx \text{ with } \Pr(\Mx=\mx)>0
\end{equation*}

In particular, if $\mx=0^p$ has positive probability, then $\X_o(0^p)=\X$, and
\begin{equation*}
Y \perp \Mx=0^p \mid \X
\end{equation*}
which is NICO.

\subsection{MARX-YO and NIMO imply MARX-YM}

Assume MARX-YO:
\begin{equation*}
\Mx=\mx \perp \X_m(\mx) \mid \X_o(\mx), Y,
\end{equation*}
and NIMO:
\begin{equation*}
\Mx=\mx \perp Y \mid \X_o(\mx).
\end{equation*}

By contraction,
\begin{equation*}
\Mx=\mx \perp \X_m(\mx), Y \mid \X_o(\mx),
\end{equation*}

which is MARX-YM.

\clearpage
\subsection{MAR does not imply MARX-YM, NIMO, or NICO}
\label{sec:MAR-MNARXYM}

We construct a counterexample showing that MAR does not imply MARX-YM, NIMO, or NICO (Table 5). Consider binary variables $X$ and $Y$, with missingness indicators $M_X$ and $M_Y$. Define the conditional distribution of $(M_X, M_Y)$ given $(X,Y)$ as in Figure~\ref{tab:MNAR-MARXYM-example}.

\begin{figure}[p]
    \centering
    \caption*{Table S1. Illustrative example of a MAR missingness scenario without MARX-YM, NICO nor NIMO}
    \begin{subfigure}[t]{0.9\textwidth}
    \caption{Fictive example with MAR but not MARX-YM nor NICO: conditional probability of $M_X, M_Y$ given $X$, $Y$ for some $a,b,c,d,e \in (0,1)$ such that all probabilities below lie in $(0,1)$\label{tab:MNAR-MARXYM-example}}
    \centering
    \begin{tabular}{cccccc}
\hline
\multirow{2}{*}{$X$} & \multirow{2}{*}{$Y$} & \multicolumn{4}{c}{$\{M_X,M_Y\}$}             \\  
                     &                        & $\{0,0\}$ & $\{0,1\}$ & $\{1,0\}$ & $\{1,1\}$ \\ \hline
$0$                    & $0$                      & $1-(a+b+d)$         & $d$ & $b$         & $a$         \\
$0$                    & $1$                      & $1-(a+c+d)$         & $d$ & $c$         & $a$         \\
$1$                    & $0$                      & $1-(a+b+e)$         & $e$ & $b$         & $a$         \\
$1$                    & $1$                      & $1-(a+c+e)$         & $e$ & $c$         & $a$         \\ \hline 
\end{tabular}
    \end{subfigure}
\vspace{1 cm}

\begin{subfigure}[t]{0.9\textwidth}
    \centering
    \caption{Resulting conditional probability table of $M_X$ given $X$, $Y$.\label{tab:MNAR-MARXYM-example2}}
    \begin{tabular}{cccc}
\hline
\multirow{2}{*}{$X$} & \multirow{2}{*}{$Y$} & \multicolumn{2}{c}{$M_X$} \\ 
                      &                        & $0$                 & $1$             \\ \hline
$0$                    & $0$                      & $1-(a+b)$           & $a+b$           \\
$0$                    & $1$                      & $1-(a+c)$           & $a+c$         \\
$1$                    & $0$                      & $1-(a+b)$           & $a+b$           \\
$1$                    & $1$                      & $1-(a+c)$           & $a+c$           \\ \hline
\vspace{0.1cm}
\end{tabular}
    \end{subfigure}
\end{figure}

By construction,
\begin{equation*}
M_X=1, M_Y=0 \perp X \mid Y
\quad
M_X=0, M_Y=1 \perp Y \mid X
\quad
M_X=1, M_Y=1 \perp Y, X
\end{equation*}

so the MAR conditions hold. However, the induced conditional distribution of $M_X$ given $(X,Y)$ (Table~\ref{tab:MNAR-MARXYM-example2}) satisfies
\begin{equation*}
M_X \not\perp (X,Y),
\quad
M_X = 0 \not\perp Y \mid X.
\end{equation*}

Therefore neither MARX-YM nor NICO hold, and therefore NIMO does not hold.

\clearpage
\section{Inconsistency of Complete Case Sub-models under General Missingness}
\label{sec:ccs-is-inconsistent}

\setcounter{figure}{0}
\renewcommand{\thefigure}{S\arabic{figure}}
\setcounter{table}{0}
\renewcommand{\thetable}{S\arabic{table}}

\subsection{Definition of Complete Case Sub-models}\label{suppl:subsub:definition-of-CCS}
Let $\mx$ be a observation pattern with $\Pr(M_Y=0,\Mx=\mx)>0$.
The Complete Case Sub-models (CCS) procedure fits, for each pattern $\mx$, a forecaster using the development subset including all the observations where the predictors observed under pattern
$\mx$ are observed (regardless of the remaining predictors).
\begin{equation*}
\Data_{\mx}
=
\left\{
(Y_i,\X_{o,i}(\mx))
:
M_{Y,i}=0,\;
\X_o(\mx) \subseteq \X_{o,i}
\right\},
\end{equation*}

Note that CCS differ from PS as CCS include, for each pattern $\m_\X$, observations which match exactly the pattern $\m_\X$ but also observations which have strictly more observed predictors. For instance, observations with all predictors observed participate in all sub-models of CCS, but only in the sub-model $\m_\X = \{0\}^p$ of PS.

\subsection{Probability limit of CCS}

Under correct model specification and standard regularity conditions,
the CCS sub-model associated with pattern $\mx$ converges to
\begin{equation*}
\pr_{\mathrm{CCS}}(\x_o,\mx)
=
\Pr\left(
Y
\mid
\X_o(\mx)=\x_o,\;
\M_o(\mx)=\mathbf{0}
\right),
\end{equation*}
where $\M_o(\mx)$ denotes the set of missingness indicators of the predictors
that are observed under pattern $\mx$.

\subsection{Comparison with MU and MC}

Recall:
\begin{equation*}
\mathrm{MU} = \Pr(Y \mid \X_o(\mx)),
\qquad
\mathrm{MC} = \Pr(Y \mid \X_o(\mx),\Mx).
\end{equation*}

In general,
\begin{equation*}
\Pr(Y \mid \X_o(\mx),\M_o(\mx)=\mathbf{0})
\neq
\Pr(Y \mid \X_o(\mx)),
\end{equation*}
and
\begin{equation*}
\Pr(Y \mid \X_o(\mx),\M_o(\mx)=\mathbf{0})
\neq
\Pr(Y \mid \X_o(\mx),\Mx=\mx)
\end{equation*}

Therefore CCS is not, in general, consistent for either MU or MC.

\subsection{Counterexample}

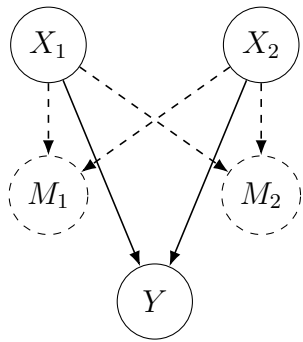
\begin{figure}[h]
    \centering
    \begin{tikzpicture}[node distance = 2cm,
state/.style={circle, draw, minimum size=1cm}]
        \node[state] (x1) {$X_1$};
        \node[state, dashed] (m1) [below of=x1] {$M_1$};
        \node[state] (y) [below right of=m1] {$Y$};
        \node[state, dashed] (m2) [above right of=y] {$M_2$};
        \node[state] (x2) [above of=m2] {$X_2$};
        
        \path[semithick] (x1) edge node {} (y);
        \path[semithick] (x2) edge node {} (y);
        \path[semithick, dashed] (x1) edge node {} (m1);
        \path[semithick, dashed] (x1) edge node {} (m2);
        \path[semithick, dashed] (x2) edge node {} (m1);
        \path[semithick, dashed] (x2) edge node {} (m2);
\end{tikzpicture}
    \caption{Illustrative scenario for unconsistency of Complete Cases Sub-models\label{fig:CCS-counterexample}}
\end{figure}

Consider $Y$ predicted from two potentially missing predictors
$X_1$ and $X_2$, with missingness indicators $M_1$ and $M_2$.
Assume the missingness mechanism depends on $(X_1,X_2)$ as shown in
Figure~\ref{fig:CCS-counterexample}.

For the pattern $(M_1=0,M_2=1)$,
CCS uses all observations with $X_1$ observed, whether or not $X_2$ is observed.
Hence the probability distribution limit is
\begin{equation*}
\Pr(Y \mid X_1, M_1=0).
\end{equation*}

However,
\begin{equation*}
\mathrm{MU}
=
\Pr(Y \mid X_1),
\end{equation*}
and
\begin{equation*}
\mathrm{MC}
=
\Pr(Y \mid X_1, M_1=0,M_2=1).
\end{equation*}

Unless additional independence conditions hold
(e.g. $M_2 \perp Y \mid X_1$),
\begin{equation*}
\Pr(Y \mid X_1,M_1=0)
\neq
\Pr(Y \mid X_1)
\quad \text{and} \quad
\Pr(Y \mid X_1,M_1=0)
\neq
\Pr(Y \mid X_1,M_1=0,M_2=1).
\end{equation*}

Thus CCS converges to a conditional distribution that differs from
both MU and MC.

\subsection{Special case: single missing predictor}

When only one predictor is subject to missingness, CCS is consistent with the MU probability distribution for the incomplete pattern, and with the MC target for the complete pattern. If additionally NICO holds, i.e.:
\begin{equation*}
Y \perp M = 0 \mid \X,
\end{equation*}

then the MC probability distribution is equal to the MU probability distribution in the complete pattern:
\begin{equation*}
\Pr(Y \mid \X,M=0)
=
\Pr(Y \mid \X),
\end{equation*}

Hence, CCS is MU-consistent, as shown by \cite{bartlett_asymptotically_2015}. In the presence of multiple missing predictors or outcome-dependent missingness, however, this equivalence no longer holds.

\clearpage
\section{Secondary and exploratory analyses}\label{suppl:secondary-analyses}

\begin{figure}[h]
  \centering
  \includegraphics[width=0.8\linewidth]{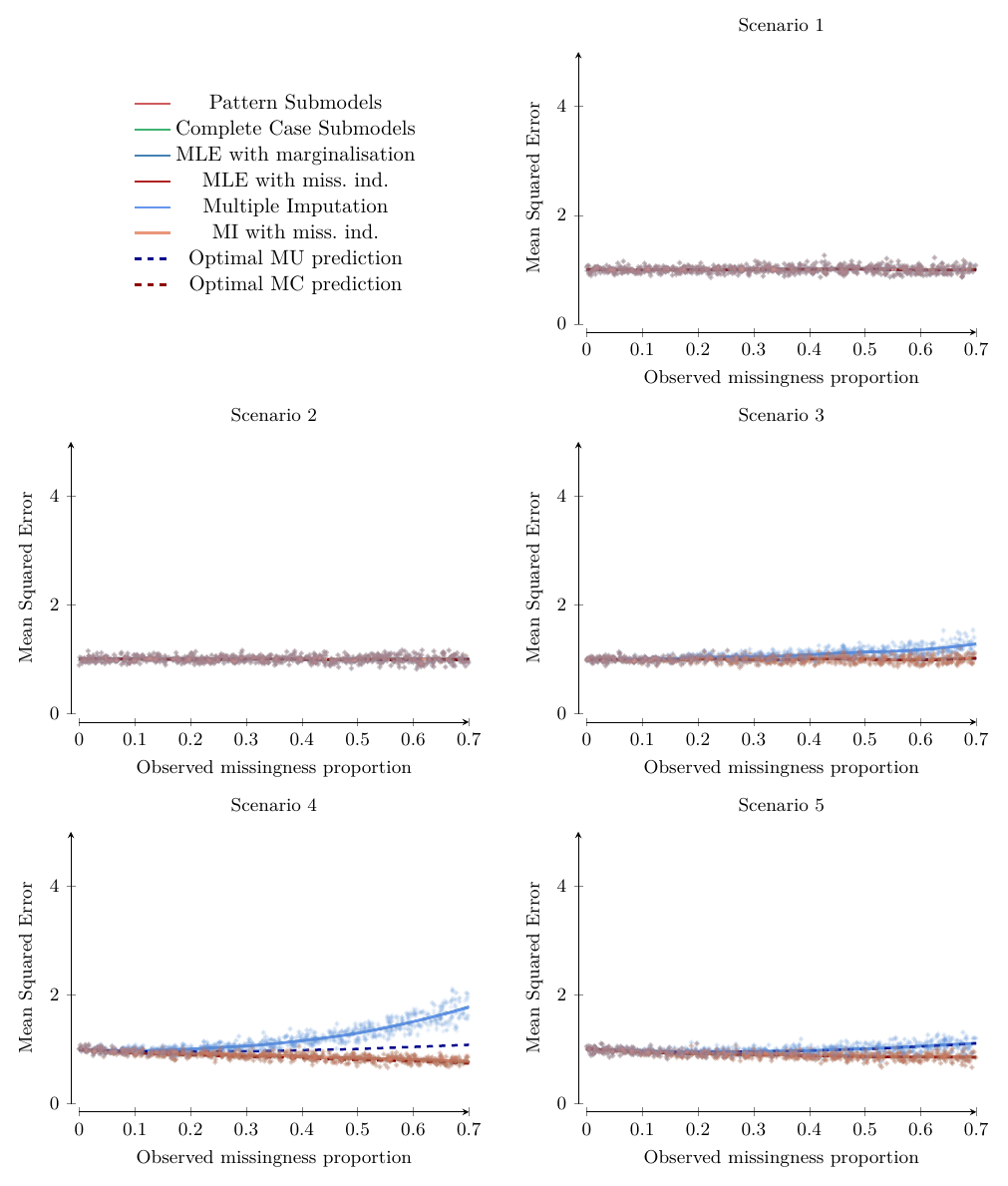}
  \caption{Mean Squared Error (MSE) for various proportions of missing values of evaluated procedures for the five scenarios, complete cases}
  \label{fig:all_scenarios_continuous_mse_complete}
\end{figure}

\begin{figure}[p]
  \centering
  \includegraphics[width=0.8\linewidth]{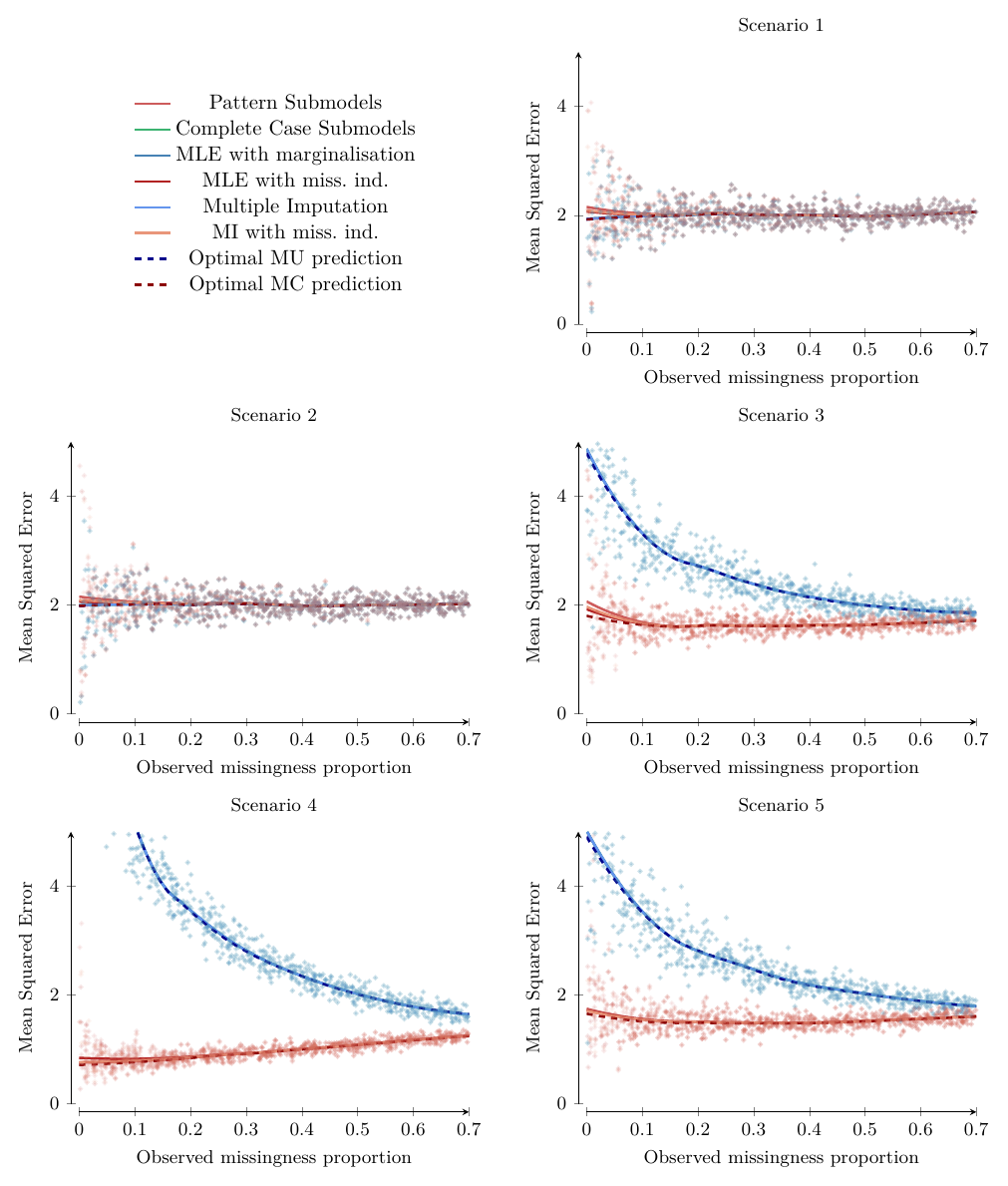}
  \caption{Mean Squared Error (MSE) for various proportions of missing values of evaluated procedures for the five scenarios, incomplete cases}
  \label{fig:all_scenarios_continuous_mse_incomplete}
\end{figure}

\begin{figure}[p]
  \centering

  \includegraphics[width=0.8\linewidth]{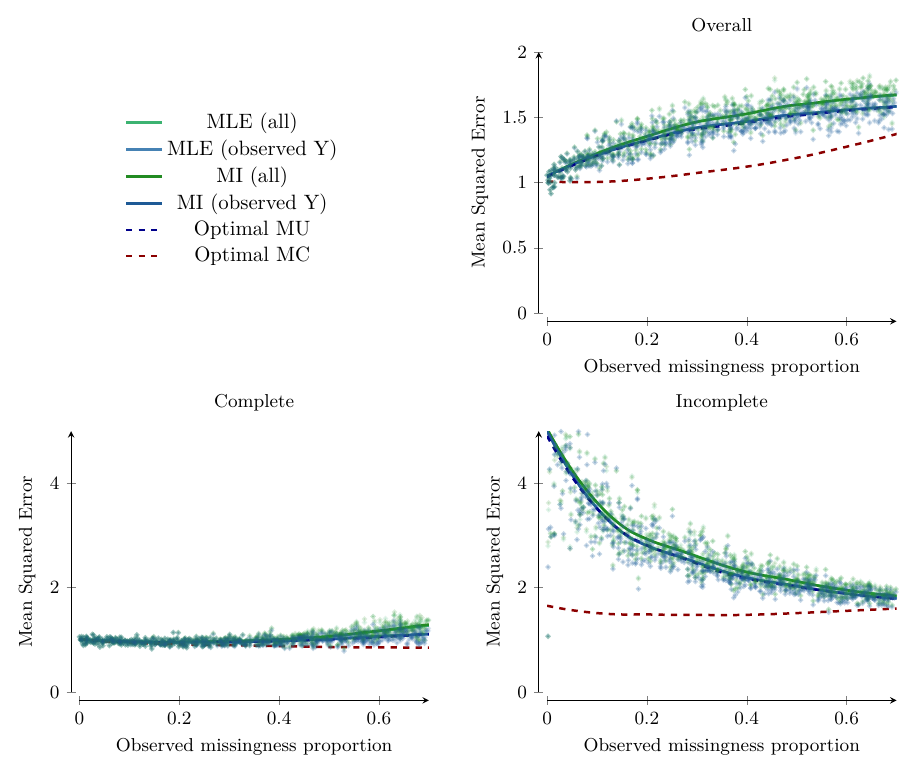}
  \caption{Mean Squared Error (MSE) for various proportions of missing values for the fifth scenario (MNAR, MARX-YO), for continuous variables, with MLE and MI methods estimated on all observations (green) or subset with observed $Y$ only (blue)}
  \label{fig:results-explo}
\end{figure}
\end{document}